%% file: main.tex
\documentclass[journal]{vgtc}                     

\onlineid{TBD}

\vgtccategory{Theoretical \& Empirical}

\title{SVLAT: Scientific Visualization Literacy Assessment Test}

\author{%
  \authororcid{Patrick Phuoc Do}{0009-0001-8966-6823},
  \authororcid{Kaiyuan Tang}{0009-0001-3512-0112},
  \authororcid{Kuangshi Ai}{0009-0005-7171-6529}, and
  \authororcid{Chaoli Wang}{0000-0002-0859-3619}
}

\authorfooter{
  \item 
  The authors are with
the Department of Computer Science and Engineering, University of Notre Dame, Notre Dame, IN 46556, USA.\\
E-mail: \{mdo23, ktang2, kai, chaoli.wang\}@nd.edu.
}

\abstract{%
\input{0_abstract}
}

\keywords{Scientific visualization, visualization literacy, assessment test, instrument, measurement, visualization education}

\graphicspath{{figs/}{figures/}{pictures/}{images/}{./}} 

\usepackage{mathptmx}
\usepackage{graphicx}
\usepackage{times}
\usepackage{algorithm}
\usepackage{algorithmic}
\usepackage{amsfonts}
\usepackage{booktabs}         
\usepackage{gensymb}
\usepackage{amsmath}
\usepackage{accessibility}
\usepackage{comment}
\usepackage{times} 
\usepackage{caption}
\usepackage{bm}
\usepackage{multirow}
\usepackage[T1]{fontenc}
\usepackage[scaled]{helvet}

\usepackage{tikz}
\usetikzlibrary{positioning,arrows.meta,shapes.geometric,calc,fit}

\usepackage{booktabs}
\usepackage{tabularx}
\usepackage{array}

\usepackage{makecell}
\usepackage{pifont}
\newcommand{\cmark}{\ding{51}}

\usepackage{xcolor}
\usepackage[dvipsnames]{xcolor}
\usepackage{anyfontsize}
\usepackage{soul}

\usepackage[table]{xcolor}
\usepackage{comment}
\usepackage{float}

\newcommand{\green}[1]{\mbox{{\color{ForestGreen} #1}}}
\newcommand{\gold}[1]{\mbox{{\color{Dandelion} #1}}}
\newcommand{\red}[1]{\mbox{{\color{Red} #1}}}

\newenvironment{myitemize}{
\begin{itemize}
 \setlength{\itemsep}{1pt}
 \setlength{\parskip}{0pt}
 \setlength{\parsep}{0pt}}{\end{itemize}
}

\begin{document}


\firstsection{Introduction}
\maketitle
\input{1_intro}

\vspace{-0.075in}
\section{Related Work}
\input{2_related}

\vspace{-0.075in}
\section{SVLAT Development Overview}
\input{3_procedure}

\vspace{-0.075in}
\section{Development of SVLAT}
\input{4_development}

\vspace{-0.075in}
\section{Discussion}
\input{5_discussion}

\vspace{-0.075in}
\section{Concluding Remarks}
\input{6_conclusion}

\vspace{-0.075in}
\section{Supplemental Materials}
\input{7_suppl}

\vspace{-0.075in}
\section{Figure Credits}
\input{8_credit}

\vspace{-0.075in}
\acknowledgments{We thank the SciVis experts, Lisa Avila, Kenneth Moreland, Hongfeng Yu, Bei Wang, and David Bauer, for participating in the SVLAT content validity evaluation. This research was supported in part by the U.S.\ National Science Foundation through grants IIS-2101696, OAC-2104158, and IIS-2401144.}

\vspace{-0.075in}
\bibliographystyle{abbrv-doi-hyperref}

\bibliography{references}

\appendix 
\crefalias{section}{appendix} 

\input{9_appendix}

\end{document}

%% file: 0_abstract.tex
Scientific visualization (SciVis) has become an essential means for exploring, understanding, and communicating complex scientific phenomena. However, the field still lacks a validated instrument assessing how well people read, understand, and interpret them. We present a scientific visualization literacy assessment test (SVLAT) that measures the general public's SciVis literacy. Covering a range of visualization forms and interpretation demands, SVLAT comprises 49 items grounded in 18 scientific visualizations and illustrations spanning eight visualization techniques and 11 tasks. Instrument development followed a staged, psychometrically grounded pipeline. We defined the construct and blueprint, followed by item generation, and expert review with five SciVis experts using the content validity ratio (mean CVR = 0.79). We subsequently administered a pilot test (30 participants) and a large-scale test tryout (485 participants) to evaluate the instrument's psychometric properties. For validation, we performed item analysis and refinement using both classical test theory (CTT) and item response theory (IRT) to examine item functioning and overall test quality. SVLAT demonstrates high reliability in the tryout sample (McDonald's $\omega_t$ = 0.82, Cronbach's $\alpha$ = 0.81). The assessment materials are available at \url{https://osf.io/hr3nw/}.

%% file: 1_intro.tex
As an indispensable means in both analysis and communication across domains ranging from business intelligence to scientific research, visualization transforms data into visuals, enabling us to perceive patterns, trends, and outliers that might be difficult to discern from raw numbers alone. Information visualization (InfoVis) is the visual representation of abstract or non-physical data. Common examples include bar charts, tables, and graphs. Scientific visualization (SciVis), in contrast, focuses on data with inherent spatial or physical structure, such as medical imaging scans, fluid dynamics simulations, or geographic models, to provide insight into the underlying phenomena. Both InfoVis and SciVis rely on the viewer's ability to interpret visual encodings and extract meaning from them. This competency is called {\bf visualization literacy}, defined as {\em the ability to read, understand, and interpret data visualizations} \cite{borner2019data, maltese2015data, beschi2025characterizing}. Visualization literacy has been recognized as a critical skill for the general public in an increasingly data-driven society \cite{cui2023adaptive}, motivating sustained work on literacy concepts, competencies, evaluation, and institutional interventions \cite{Lee-TVCG15, Firat-CGA22, Varona-arXiv25, Alper-CHI17, Chevalier-CGA18}.

Over the past decade, researchers have developed psychometrically grounded instruments for InfoVis literacy. Early work leveraged {\em item response theory} (IRT) \cite{IRT} for assessing graph comprehension \cite{Boy-TVCG14}; this line culminated in the {\em visualization literacy assessment test} (VLAT), a 53-item instrument spanning common chart types and fundamental tasks \cite{lee2016vlat}. Beyond functional reading (e.g., value extraction and pattern finding), visualization literacy is also understood to include \emph{critical judgment}, such as recognizing deceptive encodings and accounting for systematic viewer biases \cite{Camba-CGA22, Mansoor-CBV18}. Despite this progress, no standardized assessment exists for SciVis literacy, leaving core skills (3D spatial reasoning, scalar/vector interpretation, multi-field integration, and parameter awareness) unmeasured in general audiences~\cite{Wang-EVIS26}.

Meanwhile, vision–language models (VLMs) have rekindled interest in ``machine'' visualization literacy. Recent systems (e.g., GPT-4V, Gemini) can ingest visualizations and perform explanatory or question–answering tasks \cite{achiam2023gpt,team2024gemini}. Recent evaluations on visualization-literacy-style task suites report mixed results: models can succeed on some higher-level judgments (e.g., trends/extrema) yet remain brittle on basic visual evidence extraction (e.g., precise value retrieval, color-based distinctions) and can exhibit inconsistency or hallucinated readings \cite{bendeck2024empirical, li2024visualization, Bendeck-CGA25, Valentim-arXiv25}. Complementary work also investigates whether models actually rely on visual evidence (and whether they generalize under benign chart modifications), rather than answering from priors \cite{Hong-TVCG25, Dong-VIS25}. Crucially, these evaluations are InfoVis-centric because standardized, psychometrically validated SciVis instruments do not exist; as a result, rigorous human baselines and principled model benchmarking for SciVis remain underspecified~\cite{Wang-EVIS26}.

The goal of our study is to develop a standardized test to measure SciVis literacy for the general public. In doing so, we aim to complement prior InfoVis literacy assessments (like VLAT and its successors) by providing an instrument tailored to the SciVis domain. We ground our approach in established principles of psychological and educational measurement \cite{cohen2021psychological}, drawing inspiration from how earlier tests were systematically developed \cite{lee2016vlat, ge2023calvi}. In particular, our test design follows a rigorous multi-phase process, including blueprint construction, item generation, content validity, pilot test and test tryout, item analysis and test refinement, and reliability evaluation, analogous to the methodologies used for VLAT \cite{lee2016vlat}. By adhering to this process, we ensure that the resulting {\em scientific visualization literacy assessment test} (SVLAT) is both valid (covering the breadth of SciVis skills of interest) and reliable (yielding consistent results). Our contributions are as follows:
\begin{myitemize}
\vspace{-0.05in}
    \item {\bf Standardized instrument.}\ We develop SVLAT through a systematic psychometric process for assessing SciVis literacy.
    \item {\bf Validated and reliable test implementation.}\ Through an empirical study with general audiences, we show that SVLAT provides meaningful measurement of SciVis literacy. We report item difficulty and discrimination and demonstrate strong reliability.
    \item {\bf Open instrument and repository.}\ We release SVLAT by maintaining an open repository of assessment materials and visualizations to enable reproducible studies and cross-study comparisons, supporting SciVis literacy research and education.
\vspace{-0.05in}
\end{myitemize}

%% file: 2_related.tex

{\bf Visualization literacy.}\
Researchers have developed several standardized tests to measure visualization literacy, with a focus almost exclusively on InfoVis. In 2014, Boy et al. \cite{Boy-TVCG14} developed a principled IRT-based method to assess how well people can read basic graphs. They designed tests for common chart types (line graphs, bar charts, scatterplots) and demonstrated how to measure a user's latent visualization ability beyond just raw accuracy. Building on this, in 2017, Lee et al. \cite{lee2016vlat} developed VLAT, a comprehensive assessment comprising 53 multiple-choice questions (MCQs) that cover 12 types of data visualizations and eight fundamental tasks (e.g., reading values, identifying trends). To facilitate easier deployment, Pandey and Ottley recently introduced Mini-VLAT \cite{pandey2023mini}, a short-form test consisting of only 12 items that still reliably approximates a person's visualization literacy level. Mini-VLAT was validated against the original VLAT (achieving a strong correlation) and demonstrated acceptable reliability despite its brevity. Building on the success of VLAT, subsequent assessments have tackled more specialized aspects of visualization understanding. A notable example is CALVI, proposed by Ge et al. \cite{ge2023calvi} as a ``critical thinking assessment for literacy in visualizations''. This test consists of multiple-choice items that each incorporate a known chart misleader (e.g., truncated axes, inappropriate baselines, or distorted aspect ratios) and ask the viewer to interpret or judge the data. This direction aligns with broader arguments that visualization literacy should explicitly cover deception detection and bias-aware interpretation, not only routine reading tasks \cite{Camba-CGA22, Mansoor-CBV18, Firat-CGA22}. To address practical constraints of test length and adaptability, Cui et al. \cite{cui2023adaptive} developed adaptive versions of VLAT and CALVI (dubbed A-VLAT and A-CALVI, respectively), reducing assessment time while maintaining accuracy. Another recent development is the use of large language models (LLMs) to aid in item creation. For example, a technique called {\em visualization item LLM-assistance} (VILA) was used by researchers to automatically generate new test questions. Using this approach, they constructed a machine-generated VLAT variant, called VILA-VLAT \cite{cui2025promises}, which covers the same tasks and chart types as the original test. 

Despite the growing family of visualization literacy tests, it is important to note that all existing instruments focus on InfoVis. There is no established assessment of visualization literacy in the SciVis domain today. This gap is not the result of neglect, but instead arises from the strong entanglement with domain knowledge, the inherent difficulty of standardization and evaluation, a historical emphasis on algorithms over users, and limited demand from non-expert audiences.

{\bf Visualization literacy for SciVis.}\
Compared to InfoVis, SciVis places greater demands on visualization literacy for the following reasons.
First, the data are physically grounded, spatiotemporal, and often high-dimensional \cite{eden2005information, kehrer2012visualization}. 
SciVis frequently requires reconstructing 3D structures from 2D projections, managing depth and occlusion, and reasoning across time: tasks that are view- and interaction-dependent (such as navigation, slicing, and transformation). Empirical evidence on 3D volume comprehension shows that specific geometric and framing properties, such as oblique cutting planes and curved/non-planar layers, systematically hinder understanding \cite{oh2011properties}. 
Second, interpreting scientific visualizations often depends as much on domain knowledge as on general visualization literacy. 
Even in conventional chart-based visualizations, novices exhibit recurring misconceptions and evidence-seeking errors, suggesting that interpretation failures can stem from how viewers map visual evidence to claims, rather than from missing domain knowledge \cite{Rodrigues-CG21}. SciVis depictions are grounded in physical phenomena and analytic constructs, so making sense of a turbulent flow visualization or a 3D protein model presupposes familiarity with the underlying science. SciVis systems and encodings have traditionally been designed for expert analysts, implicitly assuming awareness of domain conventions, task goals, and parameter provenance (e.g., seeding strategies, isovalues, and color/opacity mappings). Without this context, non-expert viewers can misinterpret features or fail to identify what is salient, even when the depiction is perceptually clear, underscoring the importance of domain knowledge as a key prerequisite for interpretation in SciVis.
Third, SciVis often entails multi-field integration and uncertainty \cite{laramee2014future, johnson2004top}. Many applications require joint reading of multiple co-located variables (e.g., scalar, vector, and tensor fields) or across multiple modalities/time steps; fusing these facets without introducing artifacts is non-trivial and increases cognitive load for non-experts. Further, real scientific data carry measurement and model uncertainty that must be represented and interpreted (e.g., probabilistic, confidence overlays) \cite{Grigoryan-TVCG2004, Potter-Springer2011}. Decoding such composites accurately is more complicated than reading a single encoded quantity. 

Together, spatiotemporal complexity, the domain knowledge required, and the integration of multi-field and uncertainty make SciVis interpretation more demanding than InfoVis, underscoring the need for a domain-specific literacy assessment. While these challenges are central to real-world SciVis use, SVLAT in this paper intentionally targets \emph{view-independent} interpretation that can be assessed from a figure and caption under our closed-world rule. Accordingly, interaction-dependent exploration (e.g., navigation/slicing/transformation) and uncertainty-specific reasoning are treated as out of scope for this initial instrument and left to future extensions that can provide standardized interactive interfaces and uncertainty encodings.

\begin{figure*}[htb]
\centering
\includegraphics[width=0.95\textwidth]{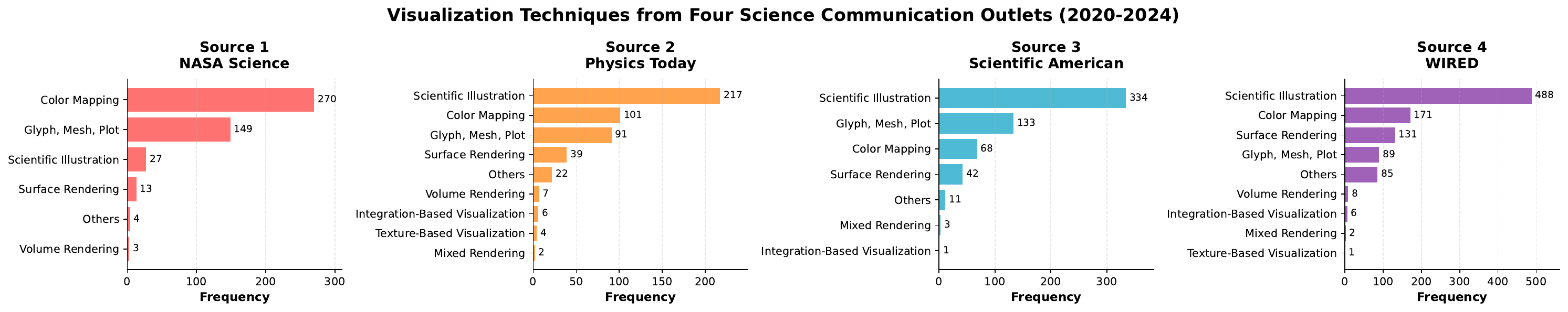}
  \vspace{-0.1in}
\caption{Observed distribution of visualization techniques in a corpus of public-facing scientific visual materials across multiple outlets. 
}
\label{fig:technique-distribution}
\end{figure*}

{\bf VLMs and ``machine'' visualization literacy.}\
The rise of VLMs, LLMs with vision capabilities, has introduced new considerations for visualization literacy in both InfoVis and SciVis. Recent VLMs (such as GPT-4 Vision \cite{achiam2023gpt} or Gemini \cite{team2024gemini}) can ingest images of data visualizations and attempt to answer questions or explain the content. Researchers have begun evaluating how well these models understand visualizations. Notably, some studies have found that state-of-the-art VLMs perform surprisingly well on standard visualization literacy tasks. For example, Li et al. \cite{li2024visualization} reported that certain VLMs can match or even outperform average humans in tasks such as identifying correlations or clusters in charts. On the other hand, more comprehensive evaluations reveal that models still have significant limitations \cite{bendeck2024empirical, pandey2025benchmarking}. 

It is worth noting that all such evaluations to date have been conducted on InfoVis, as this is the domain for which we have standardized test sets and benchmarks. In the context of SciVis, the capabilities of both humans and AI models are largely untested systematically.

%% file: 3_procedure.tex
We base our SVLAT development procedure on the principles of psychological and educational measurement \cite{cohen2021psychological} and on methodologies employed in prior developments of visualization literacy tests \cite{lee2016vlat, ge2023calvi}. In this section, we outline the high-level steps used to ensure that SVLAT is grounded in a systematic, valid design.


SVLAT was developed through a staged, psychometrically grounded pipeline that links construct definition to blueprint specification, item design, and evidence from pilot and tryout studies. This procedure mirrors the broader family of visualization literacy assessments (e.g., VLAT, Mini-VLAT, and CALVI), which similarly proceed from construct and blueprint development to expert review and large-sample tryouts, with modern psychometric analysis supporting item refinement and, in adaptive variants, efficient administration \cite{lee2016vlat, pandey2023mini, ge2023calvi, cui2023adaptive}.


We began with \textbf{construct and constraints}, defining SciVis literacy and enforcing a \emph{closed-world} interpretation rule that required each item to be answerable solely from the figure and caption. We then conducted \textbf{blueprint construction}, operationalizing the construct along two axes: (1) representative SciVis visualization techniques and (2) a SciVis-oriented task taxonomy specifying the evidence SVLAT was designed to elicit. Guided by the blueprint, \textbf{item generation} produced selected-response items, i.e., MCQs or true/false (T/F) questions with a \emph{Skip} option, along with item-specific distractors reflecting plausible interpretation errors.


Next, \textbf{content validity} involved collecting expert essentiality ratings and applying Lawshe's {\em content validity ratio} (CVR) to quantify consensus and guide item revision or removal \cite{lawshe1975quantitative}; this expert-driven screening step is also commonly used in visualization literacy test development before quantitative tryouts \cite{lee2016vlat, pandey2023mini, ge2023calvi}. We then performed a \textbf{pilot test} to assess timing, clarity, and technical presentation issues, followed by a \textbf{test tryout} to support a large-sample psychometric evaluation. Both phases were administered using Qualtrics for survey delivery and Prolific for participant recruitment. We selected Qualtrics for its flexibility in experimental design and survey flow control. Prolific was chosen to support high-quality psychometric data; prior research indicates that Prolific participants demonstrate higher attention levels, better comprehension of instructions, and lower attrition rates than participants on other crowdsourcing marketplaces, which is important for the cognitive demands of a literacy assessment \cite{peer2017beyond}.


Using tryout data, we conducted \textbf{item analysis} and \textbf{test refinement} using {\em classical test theory} (CTT) \cite{CTT} and IRT, which provide complementary perspectives on item quality. CTT offers intuitive sample-based indices, whereas IRT places items on a common latent scale and provides model-based estimates of item functioning across ability levels \cite{demars2010item, thorndike1991measurement}. For CTT, we calculated item difficulty and item discrimination using corrected item-total correlation to identify items with extreme endorsement rates or weak associations with overall test performance. For IRT, we fit a Bayesian two-parameter logistic (2PL) model to estimate item discrimination and difficulty on the latent SciVis literacy scale, quantify uncertainty through posterior summaries and credible intervals, inspect item characteristic curves (ICCs), and assess measurement precision through item and test information across ability levels \cite{Embretson-Reise00, demars2010item, Hambleton-CTTIRT93, burkner2021bayesian}. Together, these analyses guided item retention, revision, and coverage of the intended proficiency range.


Finally, \textbf{reliability evaluation} assessed the internal consistency of SVLAT using McDonald's $\omega_t$ \cite{mcdonald2013test} and Cronbach's $\alpha$ \cite{cronbach-1951}. These coefficients evaluate the consistency of item responses across the assessment and the degree to which the item set functions coherently in measuring SciVis literacy.


%% file: 4_development.tex
\begin{table*}[!t]
\centering
\setlength{\tabcolsep}{4pt}
\caption{Scientific visualization task taxonomy.}
\vspace{-0.1in}
\label{tab:task-taxonomy}
{\fontfamily{pag}\selectfont\fontsize{5.0}{6.4}\selectfont 
\begin{tabularx}{\textwidth}{@{}p{1.2cm}Xp{1.8cm}p{0.8cm}X@{}}
\toprule
\textbf{Task Family} & \textbf{Family Definition} & \textbf{Subtask} & \textbf{Abbr.} & \textbf{Subtask Definition} \\
\midrule

\multirow{2}{*}{Search} &
\multirow{2}{=}{Look for specific features in the dataset, such as checking whether a particular item is present or counting the number of occurrences.} &
Presence/Absence & P/A & Determine whether a feature is present. \\
\cmidrule(l){3-5}
 & & Counting & Cnt & Count how many instances of a feature exist. \\
\midrule

\multirow{2}{*}{\parbox[t]{1.2cm}{Pattern\\Recognition}} & 
\multirow{2}{=}{Identify trends or recurring patterns in the dataset.} &
Trend & Trnd & Detect a change or progression of a feature across the dataset. \\
\cmidrule(l){3-5}
 & & Repetition & Rep & Determine whether a pattern recurs in different regions and, if so, how often it repeats. \\
\midrule

\multirow{3}{*}{\parbox[t]{1.2cm}{Spatial\\Understanding}} &
\multirow{3}{=}{Assess positions, orientations, and spatial relationships of features.} &
Absolute & Abs & Judge the position or orientation of a feature relative to a fixed frame. \\
\cmidrule(l){3-5}
 & & Relative & Rel & Judge spatial relationships between multiple features. \\
\cmidrule(l){3-5}
 & & Intersection & Int & Determine if features intersect, occlude, or penetrate each other. \\
\midrule

\multirow{3}{*}{\parbox[t]{1.2cm}{Quantitative\\Estimation}} &
\multirow{3}{=}{Estimate and compare numeric values in the dataset.} &
Absolute Estimation & Abs & Estimate a metric in physical or data units. \\
\cmidrule(l){3-5}
 & & Relative Est (Binary) & Rel (B) & Compare two quantities using yes/no or greater/less-than judgments. \\
\cmidrule(l){3-5}
 & & Relative Est (Quant) & Rel (Q) & Compare quantities using a ratio or difference. \\
\midrule

\multirow{2}{*}{\parbox[t]{1.2cm}{Shape\\Description}} &
Characterize the geometry or topology of a structure, including form, curvature, elongation, branching, and the presence of holes. &
--- & --- & --- \\
\bottomrule
\end{tabularx}
}
\end{table*}

\subsection{Construct and Constraints}

We define the construct that SVLAT is intended to measure and the constraints necessary to preserve the intended interpretation of SVLAT outcomes. We define \textbf{SciVis literacy} as \emph{the ability to read, understand, and interpret scientific visualizations and illustrations, without any prior domain knowledge of the underlying science}. 

SVLAT targets the general public. Its results are interpreted under a {\bf closed-world rule}: performance on SVLAT items reflects the ability to interpret the figure and caption, rather than prior scientific knowledge. This rule implies that (1) any information necessary to justify the correct answer must be available within the visualization and its caption, and (2) item difficulty should arise from the interpretation of visual evidence and the caption's context, rather than from domain-specific facts.
To preserve the intended interpretation of SVLAT outcomes, SVLAT adopts the following constraints. 
First, SVLAT excludes items whose correctness depends on recalling external scientific facts, mechanisms, or specialized terminology that are not supported by the figure or caption. 
Second, SVLAT is not intended to measure the ability to create scientific visualizations or to evaluate attitudes toward science; it focuses on the ability to interpret. 
Third, item content is designed to avoid hidden prerequisites that would shift the construct toward domain expertise rather than visualization interpretation. Finally, SVLAT does not assess interaction-dependent SciVis skills (e.g., navigation, slicing, transformation) or the interpretation of explicit uncertainty visualizations, because these typically require interactive systems and additional conventions/context beyond a static figure-caption format.

\begin{figure*}[t]
  \centering
  \includegraphics[width=\textwidth]{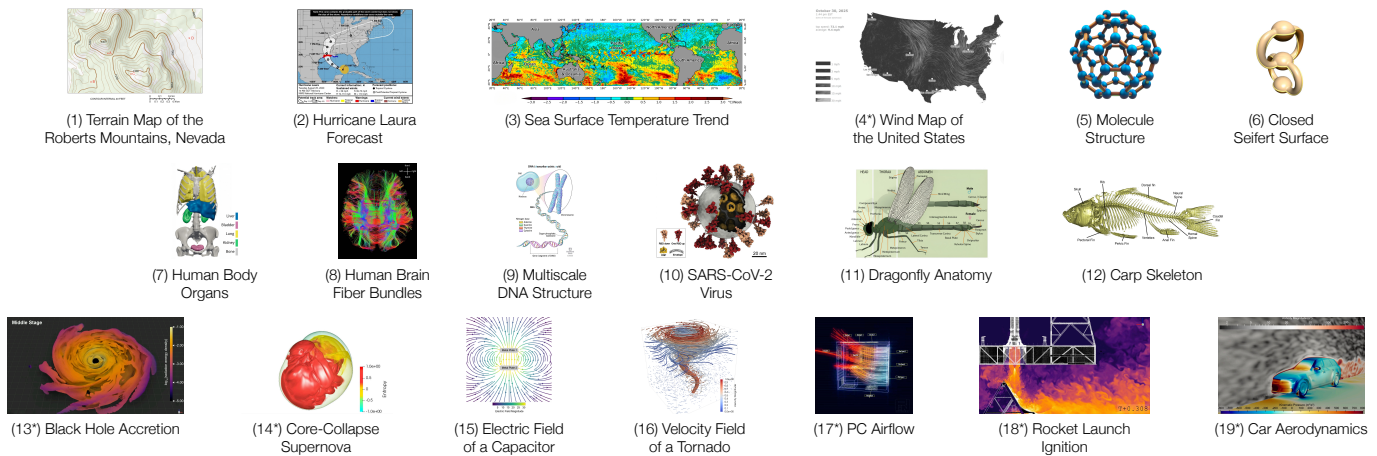}
  \vspace{-0.25in}
  \caption{The 19 visualizations considered during SVLAT development. Indices marked with * indicate animations; all others are static images. Visualization 17 is excluded based on the expert review results.} 
  \label{fig:visualizations}
\end{figure*}


\vspace{-0.05in}
\subsection{Blueprint Construction}
\label{subsec:blueprint}

To operationalize the SVLAT construct, we constructed a test blueprint that specifies SciVis techniques and visualization task types.

{\bf Visualization techniques.}\
We define the technique axis to reflect how scientific information is commonly encoded in both canonical SciVis practice and public-facing scientific materials (Figure~\ref{fig:technique-distribution}). 
First, we include technique families that are widely recognized as core SciVis techniques and are well-established in the literature: 
(1) \emph{Color Mapping} for scalar field visualization; 
(2) \emph{Volume Rendering} for volumetric scalar data and semi-transparent structure; 
(3) \emph{Surface Rendering} for boundary-focused representations such as meshes and isosurfaces; 
(4) \emph{Texture-Based Visualization} for dense visualizations of directional structure (e.g., flow patterns); 
(5) \emph{Integration-Based Visualization} for curve/surface constructions driven by integration through vector fields (e.g., streamlines and related flow representations); and 
(6) \emph{Mixed Rendering} that combines complementary visualizations to support interpretation under occlusion and multi-attribute conditions \cite{telea2014data, wright2007introduction}. 
Second, we include 
(7) \emph{Glyph, Mesh, Plot} as a family of symbolic and abstracted representations. This categorization is theoretically grounded in Tory and Möller's taxonomy \cite{Tory-InfoVis04}, which classifies these techniques under discrete design models. Glyphs are commonly used in SciVis to encode local attributes (e.g., direction and magnitude) \cite{telea2014data}. Meshes rely on an underlying discrete topology of nodes and connections. Plots, though often associated with InfoVis, function as SciVis when the spatialization is ``given'' (inherent to the data) rather than chosen by the designer \cite{Tory-InfoVis04}. Interpreting this family requires a specific literacy in discrete mark semantics and relational structures.
Third, we include 
(8) \emph{Scientific Illustration} to capture schematic or explanatory depictions that rely on diagrammatic conventions; SciVis research and illustrative visualization work explicitly motivates illustration-inspired abstraction and emphasis to improve interpretability of complex scientific visuals, and scientific publishing practice treats conceptual/summary figures as standard communication artifacts with an emphasis on clarity and accessibility for broad audiences \cite{gooch2005illustrative}.

{\bf Visualization tasks.}\ 
We define the task axis using a SciVis-oriented interpretation taxonomy that draws from domain-independent classification for volume analysis and model-based visualization theory \cite{laha2015classification, Tory-InfoVis04}. Specifically, SVLAT adopts the five high-level categories proposed by Laha et al., who defined these goal-based tasks as ``work operators'' required for accomplishing a main scientific objective \cite{laha2015classification}. This framing enables SVLAT to isolate the cognitive work of interpretation from the ``enabling'' tasks of tool interaction, which are outside the scope of this assessment \cite{laha2015classification}. This taxonomy is also theoretically supported by the high-level classification, which identifies specific tasks enabled by ``given'' spatialization and ``continuous'' design models \cite{Tory-InfoVis04} inherent to the SciVis domain. 
The taxonomy comprises five high-level categories. 
\emph{Search} tasks assess whether participants can locate and identify information in a visualization, operationalized as \emph{Presence/Absence} and \emph{Counting}. 
\emph{Pattern Recognition} tasks evaluate the ability to detect regularities in the displayed data, operationalized as \emph{Trend} and \emph{Repetition}. 
\emph{Spatial Understanding} tasks target reasoning about positions and spatial relationships, operationalized as \emph{Absolute}, \emph{Relative}, and \emph{Intersection}. 
\emph{Quantitative Estimation} tasks capture value extraction and comparison from visual encodings, operationalized as \emph{Absolute Estimation}, \emph{Relative Estimation (Binary)}, and \emph{Relative Estimation (Quantitative)}. 
Finally, \emph{Shape Description} tasks assess the ability to characterize geometric or structural properties based on what is shown, operationalized as \emph{Shape Description}. 
Together, these operationalizations yield 11 task types, as outlined in Table~\ref{tab:task-taxonomy}. This granularity supports targeted item authoring to ensure balanced coverage and enables subsequent task-level diagnostics during refinement and validation.

\vspace{-0.075in}
\subsection{Visualization Preparation}

\subsubsection{Coverage}
\label{subsec:coverage}

We prepared the SVLAT visualization set to (1) cover the eight visualization techniques described in Section~\ref{subsec:blueprint} and avoid overfitting SVLAT to a narrow data regime. To this end, we intentionally diversified the set across (2) \emph{data fields}, (3) \emph{data dimensionalities}, and (4) \emph{application domains}. The resulting collection encompasses scalar, vector, tensor, and mixed-field representations, spanning 2D, 3D, and temporal (time-varying) settings. We also sought breadth across scientific domains (including astronomy, biology, chemistry, earth system science, mathematics, medical science, and physics) to reduce dependence on familiarity with any single discipline. Finally, the set includes both \emph{static images} and \emph{animations}: techniques that rely on temporal change are represented as short video clips, whereas other techniques are represented as still images, enabling SVLAT to assess interpretation in both static and dynamic SciVis contexts.


\vspace{-0.075in}
\subsubsection{Visualization Retrieval}
\label{subsec:retrieval}

SVLAT visualizations were retrieved from multiple source categories to balance \emph{authority}, \emph{diversity}, and \emph{public familiarity}. 
The 19 visualizations chosen from a collection of 51 examples are shown in Figure~\ref{fig:visualizations}. 
Quality control was conducted to ensure that images and videos have sufficient resolution, high-quality rendering, and clearly visible legends.

{\bf Official government science agencies.}\
We included visualizations from official U.S.\ government science organizations (e.g., NOAA/NHC and USGS for earth system science, and NASA for astronomy). These sources provide high-quality, curated visual materials with stable provenance and standardized conventions that many readers have encountered in public-facing science communication.

{\bf Academic and institutional sources.}\
Some visualizations were created by universities, institutions, and research laboratories. These sources provide high-quality educational figures and domain-grounded representations that remain accessible to non-specialists.

{\bf Peer-reviewed research papers.}\
We included visualizations from academic publications to represent canonical SciVis techniques and research-grade examples. When adapting paper figures, we prioritized \emph{simple} single-view visualizations and avoided dense annotation and visualization layouts that depend on extensive context.

{\bf Science journalism and educational websites.}\
We incorporated visualizations from science magazines and educational websites, often curated for broad audiences and emphasizing clarity of communication.

{\bf Online platforms and user-generated videos.}\
To broaden the space of dynamic visualization and modern visualization styles, we also included short clips from online platforms. These sources frequently present visualizations commonly encountered by the general public and employ animated or simulation-based explanatory styles.

{\bf In-house rendering using ParaView.}\
For several scanned and simulation datasets, we generated visualizations ourselves using ParaView. The main benefits are direct control, output quality, and reproducibility. ParaView enables precise configuration of the visualization pipeline and rendering parameters (e.g., views, transfer functions, and annotations), and supports exporting high-resolution visualizations.

{\bf Rationale for source diversity.}\
Using multiple source categories reduces dependence on any single visual convention or distribution channel, helping to keep the benchmark close to what a general audience may encounter in practice: a mixture of official reports, educational/institutional figures, research-derived imagery, and modern web-based animated content. This diversity also strengthens the ecological validity of SVLAT while retaining clear provenance and attribution.

\begin{table*}[t]
\centering
\caption{The SVLAT blueprint (19 visualizations, 71 items). A \cmark\ denotes at least one item instantiating the technique-task pairing; blank cells indicate pairings not sampled in this version (not necessarily infeasible), reflecting test-length and construct constraints. 
}
\vspace{-0.1in}
\label{tab:vis-tech-tasks}
\setlength{\tabcolsep}{4pt}
{\fontfamily{pag}\selectfont\fontsize{5.0}{6.4}\selectfont 
\begin{tabular}{@{}>{\centering\arraybackslash}m{2cm}|
                >{\centering\arraybackslash}m{1.2cm}|cc|cc|ccc|ccc|c@{}}
\hline
\multirow{3}{*}{\makecell[c]{\textbf{Visualization}\\\textbf{Technique}}} &
\multirow{3}{*}{\makecell[c]{\textbf{Visualization}\\\textbf{ID}\\\textbf{(Figure~\ref{fig:visualizations})}}} &
\multicolumn{11}{c}{\textbf{Task}} \\
\cline{3-13}
& &
\multicolumn{2}{c}{\textbf{Search}}
& \multicolumn{2}{c}{\textbf{Pattern Recognition}}
& \multicolumn{3}{c}{\textbf{Spatial Understanding}}
& \multicolumn{3}{c}{\textbf{Quantitative Estimation}}
& \textbf{Shape Description} \\
\cline{3-13}
& &
\makecell[c]{Presence/Absence}
& \makecell[c]{Counting}
& \makecell[c]{Trend}
& \makecell[c]{Repetition}
& \makecell[c]{Absolute}
& \makecell[c]{Relative}
& \makecell[c]{Intersection}
& \makecell[c]{Absolute\\Estimation}
& \makecell[c]{Relative Est\\(Binary)}
& \makecell[c]{Relative Est\\(Quant)}
& \\
\hline

Color Mapping                    & 3 & \cmark &  & \cmark &  &  &  &  & \cmark & \cmark &  &  \\ 
Volume Rendering             & 12, 14*, 18* &        &  & \cmark &  &  & \cmark &  &  & \cmark &  &  \\ 
Surface Rendering            & 6, 7, 13* &        & \cmark & \cmark &  & \cmark & \cmark & \cmark &  & \cmark &  & \cmark \\ 
Texture-Based Vis.  & 4* &        &  &  & \cmark &  &  &  &  & \cmark &  & \cmark \\ 
Integration-Based Vis.
                             & 8, 15, 17* &        &  & \cmark & \cmark  &  &  & \cmark  &  & \cmark &  &  \\ 
Mixed Rendering              & 16, 19* &        &  & \cmark &  & \cmark &  &  &  & \cmark & \cmark & \cmark \\ 
Glyph, Mesh, Plot           & 1, 2, 5 & \cmark & \cmark & \cmark & \cmark & \cmark & & \cmark & \cmark & \cmark & \cmark & \cmark\\ 
Scientific Illustration      & 9, 10, 11 & \cmark & \cmark &  &  & \cmark & \cmark &  & \cmark &  & \cmark &  \\ \hline

\end{tabular}
}
\end{table*}


\vspace{-0.075in}
\subsubsection{Caption Generation}
\label{subsec:captions}

Captions were designed to be \emph{complementary} to the visualization: they provide essential context without giving away answers. 
When available, we used the visualization's source description to draft an initial caption that accurately states (1) the phenomenon or data being shown, (2) the visualization technique and its primary encodings, and (3) any necessary units, spatial/temporal scope, or viewpoint assumptions. 
We then refined each caption for a general audience by preserving the source's factual claims and including attribution, explicitly describing the primary encodings (e.g., ``{\em color indicates temperature}'') while avoiding unexplained abbreviations or jargon, keeping the text short enough to read quickly during testing, and avoiding wording that would trivially reveal the correct response for any associated item.
Finally, we leveraged the ChatGPT 5.1 Thinking model as a \emph{revision assistant} to improve readability and audience accessibility. We prompted the model to propose multiple alternatives for each caption (e.g., \emph{clear and concise}, \emph{slightly more formal}, and \emph{alternative for clarity}), and then manually selected and verified the final wording to ensure correctness and alignment with the visualization.

\vspace{-0.075in}
\subsection{Item Generation}

Guided by the SVLAT blueprint, we generated an initial pool of items by instantiating task-technique combinations with the prepared visualization and drafting questions to elicit the intended evidence targets. Each item was designed to assess \emph{one} primary task type from the task taxonomy to support clean blueprint coverage accounting and to enable task-level diagnostics during later refinement.

{\bf Item format.}\
SVLAT items use selected-response formats (MCQ or T/F) and include a \emph{Skip} option. 
We excluded the open-ended format as it is difficult to evaluate reliably via judgmental evaluation and qualitative analysis \cite{lee2016vlat}. 
The primary format is MCQ because it supports objective scoring, efficient administration at scale, and item diagnostics through response-option patterns. Excluding the skip option, most MCQs contain four options, with two using three; the remaining items are in T/F format. The skip option is included to reduce forced guessing; we report item-level skip rates in the analysis. 
T/F items were used when an MCQ formulation would otherwise require long, sentence-heavy alternatives (e.g., ``{\em Which statement most accurately describes \dots}''), which can increase reading burden and shift item difficulty toward verbal processing rather than visualization interpretation. In these cases, we reformulated the item into concise evaluative statements. 

{\bf Drafting and revision.}\
We iteratively drafted and revised items, utilizing ChatGPT 5.1 to enhance clarity and maintain alignment with SVLAT's construct and closed-world interpretation rule.
Revisions were applied to both captions and questions across the item pool, with the goal that each item could be answered using figure-and-caption evidence alone and that difficulty arose from visualization interpretation rather than from domain knowledge or wording complexity.

{\bf Distractor construction.}\
For MCQs, distractors were written to be plausible and construct-relevant, reflecting common misinterpretations of the visualization (e.g., selecting an incorrect region; alternating or reversing an ordering implied by a legend or scale; or confusing absolute vs.\ relative judgments). Distractors were constructed item-specifically rather than using fixed templates, and we avoided distractors that rely primarily on subtle wording.

{\bf Primary-task assignment and adjudication.}\
To ensure that each item was assigned to exactly one primary task type, three researchers independently assigned a task label to each drafted item. Disagreements were resolved through discussion until a consensus was reached. Most conflicts arose for items involving finding extrema (e.g., identifying the maximum/minimum). We categorized these items as \emph{Quantitative Estimation - Relative Estimation (Binary)} because the dominant cognitive operation is comparative: identifying an extremum typically entails repeated pairwise comparisons to determine whether one candidate is larger/smaller than another. Other disagreements typically arose when an item could plausibly be mapped to multiple task types (e.g., requiring both a search and a comparison step). In such cases, we resolved labels by selecting a single \emph{primary} task that best reflected the item's intended evidence target and primary source of difficulty, treating any additional operations as secondary. After consensus labeling, only disputed items were reviewed and confirmed by a SciVis expert. 
Ultimately, this process resulted in an SVLAT item bank comprising 71 items. Accordingly, the blueprint is presented in Table~\ref{tab:vis-tech-tasks}, which balances visualization techniques and tasks.

\vspace{-0.05in}
\subsection{Ethical Considerations}

All human-subject procedures conducted, including the expert review and crowd-sourced testing, were approved by the Institutional Review Board at the University of Notre Dame (No.\ 18-01-4334). Before accessing the study materials, informed consent was obtained from all experts and crowd-workers. Participants were provided with a clear description of the study goals, procedures, and data-use policies, and they indicated their consent by completing an electronic consent form.

\vspace{-0.05in}
\subsection{Content Validity}

Before administering SVLAT to a general audience, we assessed the content validity of the drafted items through an expert review. We recruited a panel of five SciVis experts, comprising two academics, two industry professionals, and one representative from a national lab. The panel included experts across a range of ages (25 to 64) and with extensive SciVis experience (6 to 20+ years). The goal of this step was to verify that each item is relevant to the SVLAT construct and that the item pool collectively provides appropriate coverage of the blueprint.

{\bf Rating procedure.}\
Before the review, we informed experts of the evaluation's purpose, provided our definition of SciVis literacy, and shared the SVLAT task taxonomy. Experts were provided with the full item list (71 items), each comprising the visualization and caption, the question stem with response options, and the intended primary task label from the taxonomy. 
Experts independently evaluated each item using Lawshe's three-level essentiality judgment: \emph{essential}, \emph{useful but not essential}, or \emph{not necessary} \cite{lawshe1975quantitative} for measuring SciVis literacy. Ratings were collected for all items, and experts could optionally provide brief written comments when they perceived ambiguity, construct mismatch, reliance on domain knowledge, or unclear wording/captioning.

{\bf Optional visualization-level feedback.}\
In addition to the essentiality ratings used for CVR, we collected optional free-text feedback to improve the clarity and accessibility of the visualizations for the general public. For each visualization, experts could provide comments and suggestions (e.g., color choice, label size, visual clutter, and overall clarity for non-expert participants). We also included two optional open-ended prompts at the end of the review: (1) \emph{Exclusion} where experts could recommend specific visualization(s) to remove and briefly justify the recommendation (e.g., redundancy, limited representativeness, or excessive difficulty for the target audience); and (2) \emph{Inclusion} where experts could suggest additional scientific visualization(s) for future SVLAT versions, including the intended skills assessed and a rationale for inclusion. These qualitative inputs were not used in CVR computation. However, they informed subsequent revisions to the visualizations and captions and guided decisions about representativeness and future refinement of the visualization set.

{\bf CVR computation and decision rules.}\
For each item, we computed the CVR following Lawshe \cite{lawshe1975quantitative}: $\mathrm{CVR} = (n_e - N/2)/(N/2)$, where $N$ is the number of experts and $n_e$ is the number of experts who rated the item as \emph{essential}. With $N=5$, CVR values range from $-1$ to $1$, where larger values indicate a stronger consensus that an item is essential. We used the CVR values as an initial screening signal and excluded items with $\mathrm{CVR}<0$, indicating that fewer than half of the experts rated the item as essential. Items at or above this threshold were retained for further refinement and piloting.

{\bf Post-review refinement.}\
After the CVR computation for all candidate items, we conducted a post-review refinement to finalize the item pool by synthesizing expert feedback and quantitative outcomes. 
Based on Lawshe's decision rules, we excluded items with a CVR below zero, indicating that more than half of the experts did not reach consensus on their essentiality for measuring SciVis literacy. This led to the removal of five items (8, 11, 21, 42, 44).
We also audited the pool for redundancy within each visualization technique. We identified four items (26, 48, 50, 53) that were near-duplicates of retained items in visualization evidence and primary task demand. Because these items added little new construct coverage and received only marginal essentiality support (CVR = 0.2), we removed them to reduce participant burden and prevent further over-representation of the same technique-task pairings, while maintaining coverage consistent with the blueprint. 

In addition, despite initial quality controls, we removed the PC Airflow visualization and its three associated items (63, 64, 65) after experts identified critical legibility and presentation hurdles that emerged during high-resolution or full-screen viewing. Specifically, experts noted that essential labels became blurred over the animation and were difficult to distinguish. These refinements were crucial in upholding our closed-world rule, ensuring that item difficulty arises solely from the cognitive interpretation of visual evidence, rather than from technical artifacts or presentation-based limitations. 
To improve the assessment's rigor, we refined the wording of ambiguities identified by experts to ensure that each question's intended task demand precisely matched its taxonomy label and that the difficulty arose from visualization interpretation rather than linguistic complexity. 
After this stage, the retained SVLAT item bank comprises 59 items with an average CVR of 0.78.

\begin{table*}[!t]
\centering
\setlength{\tabcolsep}{4pt}
\caption{The list of 51 SVLAT tryout items and their corresponding primary task, content validity ratio (CVR), CTT item difficulty ($P$) and item discrimination  ($r$), and IRT item easiness ($e$) and item discrimination ($a$). In the CTT columns, \green{green} indicates easy items or high discrimination, \gold{yellow} indicates moderate difficulty or medium discrimination, and \red{red} indicates hard items or low discrimination. Items 49 and 69 are excluded from the final set. 
}
\vspace{-0.1in}
\label{tab:item-list}
{\fontfamily{pag}\selectfont\selectfont\fontsize{5.0}{6.4}\selectfont
\begin{tabularx}{\textwidth}{@{}>{\raggedright\arraybackslash}m{2.0cm}%
>{\raggedright\arraybackslash}p{0.7cm}%
>{\raggedright\arraybackslash}X%
>{\raggedright\arraybackslash}p{1.6cm}%
>{\centering\arraybackslash}p{0.4cm}%
>{\centering\arraybackslash}p{0.4cm}%
>{\centering\arraybackslash}p{0.4cm}%
>{\centering\arraybackslash}p{0.4cm}%
>{\centering\arraybackslash}p{0.4cm}@{}}
\toprule
\multirow{2}{*}{\textbf{Visualization}} & 
\multirow{2}{*}{\textbf{Item ID}} & 
\multirow{2}{*}{\textbf{Stem}} & 
\multirow{2}{*}{\textbf{Task}} & 
\multirow{2}{*}{\textbf{CVR}} & 
\multicolumn{2}{c}{\textbf{CTT}} & 
\multicolumn{2}{c}{\textbf{IRT}} \\
\cmidrule(lr){6-7} \cmidrule(lr){8-9}
& & & & & \textbf{$P$} & \textbf{$r$} & \textbf{$e$} & \textbf{$a$} \\
\midrule

\multirow[c]{5}{2.7cm}{\makecell[l]{Terrain Map of the}\\{Roberts Mountains,}\\{Nevada}}
& Item 1  & What is the elevation of waypoint D? & Quant Est - Abs & 1   & \gold{0.71} & \gold{0.25} & 0.04  & 0.66 \\
& Item 2  & Which waypoint has the highest elevation? & Quant Est - Rel (B) & 1   & \gold{0.84} & \green{0.41} & 0.14  & 1.38 \\
& Item 3  & By approximately how many feet does the elevation of waypoint A exceed that of waypoint D? & Quant Est - Rel (Q) & 0.6 & \red{0.39} & \gold{0.25} & -2.24 & 0.65 \\
& Item 4  & Moving from waypoint C to waypoint A along a straight line, the \textbf{elevation} \_\_\_\_ . & Pattern Rec - Trnd & 1   & \gold{0.56} & \gold{0.23} & -0.95 & 0.50 \\
& Item 5  & Moving from waypoint C to waypoint B along a straight line, the \textbf{slope} is \_\_\_\_ . & Pattern Rec - Trnd & 0.6 & \gold{0.61} & \gold{0.25} & -0.68 & 0.69 \\
\midrule

\multirow[c]{4}{2.7cm}{\makecell[l]{Hurricane Laura}\\{Forecast}}
& Item 7  & From 7 AM Wed to 7 AM Sun, sustained winds do not exceed 110 mph anywhere along the forecast track. & Search - P/A & 1   & \gold{0.75} & \green{0.31} & -0.16 & 1.03 \\
& Item 9  & From 10 AM Tue to 7 PM Wed, in which direction does the hurricane primarily move? & Spatial Und - Abs & 0.6 & \green{0.88} & \gold{0.29} & 0.70  & 1.11 \\
& Item 10 & The hurricane is expected to reach the coast between \_\_\_\_. & Spatial Und - Abs & 0.6 & \green{0.88} & \green{0.35} & 0.25  & 1.58 \\
& Item 12 & From 10 AM Tue to 7 AM Fri, the hurricane's sustained wind speed \_\_\_\_ . & Pattern Rec - Trnd & 1   & \gold{0.67} & \green{0.34} & -0.64 & 1.07 \\
\midrule

\multirow[c]{5}{2.7cm}{\makecell[l]{Sea Surface}\\{Temperature Trend}}
& Item 13 & During the past week, the sea surface temperatures in the Pacific Ocean between 0-20°N and 140-160°E mostly \_\_\_\_ . & Quant Est - Abs & 1   & \gold{0.72} & \green{0.43} & -0.48 & 1.23 \\
& Item 14 & Which ocean region has cooled the most over the past week? & Quant Est - Rel (B) & 0.2 & \gold{0.57} & \gold{0.11} & -0.68 & 0.37 \\
& Item 15 & At least one region of the Atlantic Ocean experiences strong warming (greater than 2°C per week). & Search - P/A & 1   & \gold{0.78} & \red{0.02} & 1.96  & 0.37 \\
& Item 16 & What are the trends in sea surface temperatures across the Pacific Ocean over the past week? & Pattern Rec - Trnd & 1   & \green{0.91} & \green{0.30} & 0.68  & 1.43 \\
\midrule

\multirow[c]{4}{2.7cm}{\makecell[l]{Wind Map of the}\\{United States}}
& Item 17 & Which statement most accurately describes the overall wind pattern? & Pattern Rec - Trnd & 1   & \gold{0.83} & \gold{0.24} & 0.85  & 0.76 \\
& Item 18 & Which city has the highest wind speed? & Quant Est - Rel (B) & 0.6 & \gold{0.79} & \green{0.42} & -0.16 & 1.46 \\
& Item 19 & In the region between Columbus and New York, the winds \_\_\_\_ . & Shape Description & 0.2 & \gold{0.84} & \gold{0.22} & 0.72  & 0.86 \\
& Item 20 & The winds near Chicago and Houston flow in roughly the same direction. & Pattern Rec - Rep & 1  & \gold{0.83} & \green{0.35} & 0.33  & 1.13 \\
\midrule

\multirow[c]{2}{2.7cm}{\makecell[l]{Molecule Structure}}
& Item 22 & How many other atoms is each atom directly connected to? & Search - Cnt & 0.6 & \gold{0.84} & \green{0.35} & 0.22  & 1.30 \\
& Item 23 & Which structural pattern can be observed in the molecule? & Pattern Rec - Rep & 0.2 & \gold{0.65} & \green{0.39} & -0.72 & 1.06 \\
\midrule

\multirow[c]{5}{2.7cm}{\makecell[l]{Closed Seifert Surface}}
& Item 25 & How many holes does the object have? & Search - Cnt & 0.6 & \green{0.89} & \gold{0.13} & 1.71  & 0.73 \\
& Item 27 & Suppose you make a single cut through the curved tube halfway between the top and middle balls. What happens to the balls after the cut? & Shape Description & 0.6 & \gold{0.72} & \green{0.34} & -0.25 & 0.99 \\
& Item 28 & It is possible to travel from point A to point B by moving only along the light brown surface, without going through empty space or onto any dark-brown region or light-yellow ridge. & Spatial Und - Int & 1   & \green{0.88} & \gold{0.27} & 0.82  & 1.04 \\
\midrule

\multirow[c]{2}{2.7cm}{\makecell[l]{Human Body Organs}}
& Item 31 & From this viewpoint, which organ is closest to the viewer? & Spatial Und - Rel & 1   & \gold{0.73} & \gold{0.13} & 1.04  & 0.42 \\
& Item 32 & Which pair of organs is the farthest apart? & Spatial Und - Rel & 1   & \green{0.88} & \gold{0.19} & 1.42  & 0.79 \\
\midrule

\multirow[c]{2}{2.7cm}{\makecell[l]{Human Brain}\\{Fiber Bundles}}
& Item 33 & The fiber bundles within the marked region generally run front-back. & Pattern Rec - Trnd & 0.6 & \green{0.86} & \green{0.37} & 0.34  & 1.29 \\
& Item 35 & Which region has the highest density of fiber bundles? & Quant Est - Rel (B) & 1   & \green{0.88} & \gold{0.25} & 0.97  & 0.98 \\
\midrule

\multirow[c]{2}{2.7cm}{\makecell[l]{Multiscale}\\{DNA Structure}}
& Item 36 & Relative to the two sugar-phosphate backbones, the nitrogen base pairs are located \_\_\_\_. & Spatial Und - Rel & 0.6 & \green{0.90} & \gold{0.23} & 1.11  & 1.01 \\
& Item 37 & Which of the following correctly lists these structures from largest to smallest? & Spatial Und - Rel & 1   & \gold{0.72} & \green{0.33} & -0.15 & 0.87 \\
\midrule

\multirow[c]{2}{2.7cm}{\makecell[l]{SARS-CoV-2 Virus}}
& Item 40 & Starting from the center of the particle and moving outward, which order is correct? & Spatial Und - Rel & 1   & \gold{0.79} & \green{0.40} & -0.05 & 1.25 \\
& Item 41 & What is the approximate radius of the lipid envelope? & Quant Est - Abs & 1   & \red{0.43} & \gold{0.15} & -2.18 & 0.40 \\
\midrule

\multirow[c]{3}{2.7cm}{\makecell[l]{Dragonfly Anatomy}}
& Item 43 & How many segments are present in the abdomen of a \textbf{female} dragonfly? & Search - Cnt & 0.6 & \gold{0.84} & \gold{0.28} & 0.52  & 1.03 \\
& Item 45 & Which of the following parts is present only in a \textbf{male} dragonfly? & Search - P/A & 0.6 & \green{0.90} & \gold{0.12} & 1.67  & 0.79 \\
& Item 46 & Which part of the body are the legs attached to? & Spatial Und - Abs & 0.6 & \green{0.91} & \gold{0.13} & 2.05  & 0.70 \\
\midrule

\multirow[c]{2}{2.7cm}{\makecell[l]{Carp Skeleton}}
& Item 47 & Fin rays are the thin bones that support each fin. In which labeled fin are the fin rays the longest? & Quant Est - Rel (B) & 0.6 & \gold{0.70} & \gold{0.12} & 0.64  & 0.43 \\
& Item 49 & \textcolor{gray}{How are the ribs arranged in relation to the vertebrae?} & Spatial Und - Rel & 0.6 & \red{0.18} & \red{0.06} & -4.72 & 0.47 \\
\midrule

\multirow[c]{2}{2.7cm}{\makecell[l]{Black Hole Accretion}}
& Item 51 & Over the course of the animation, how does the shape of the gas change? & Pattern Rec - Trnd & 0.6 & \green{0.87} & \gold{0.23} & 0.95  & 0.91 \\
& Item 52 & In the middle stage, which region has the highest energy? & Quant Est - Rel (B) & 1   & \gold{0.63} & \gold{0.16} & -0.05 & 0.41 \\
\midrule

\multirow[c]{3}{2.7cm}{\makecell[l]{Core-Collapse}\\{Supernova}}
& Item 54 & Over the course of the animation, the red internal gas \_\_\_\_ . & Pattern Rec - Trnd & 0.6 & \gold{0.52} & \gold{0.23} & -1.24 & 0.55 \\
& Item 55 & The internal shocked gas always has lower entropy than the stalled shockwave. & Quant Est - Rel (B) & 1   & \gold{0.66} & \gold{0.23} & -0.21 & 0.59 \\
& Item 56 & As the red internal gas expands on one side, the outer shell on that side \_\_\_\_ . & Pattern Rec - Trnd & 0.6 & \gold{0.72} & \gold{0.29} & -0.04 & 0.77 \\
\midrule

\multirow[c]{2}{2.7cm}{\makecell[l]{Electric Field}\\{of a Capacitor}}
& Item 57 & Where is the electric field magnitude the highest? & Quant Est - Rel (B) & 1   & \green{0.97} & \red{0.04} & 2.95  & 0.84 \\
& Item 59 & Which statement most accurately describes the electric field lines? & Spatial Und - Rel & 0.6 & \gold{0.83} & \gold{0.23} & 0.71  & 0.82 \\
\midrule

\multirow[c]{3}{2.7cm}{\makecell[l]{Velocity Field}\\{of a Tornado}}
& Item 60 & In a top-down view (looking down from above), the tornado rotates clockwise. & Spatial Und - Abs & 1   & \green{0.87} & \gold{0.24} & 1.18  & 0.82 \\
& Item 61 & The arrangement of the streamlines forms a 3D structure best described as \_\_\_\_ . & Shape Description & 0.6 & \gold{0.66} & \gold{0.18} & -0.06 & 0.53 \\
& Item 62 & In which region is the tornado's velocity magnitude the highest? & Quant Est - Rel (B) & 1   & \gold{0.54} & \gold{0.22} & -1.07 & 0.53 \\
\midrule

\multirow[c]{2}{2.7cm}{\makecell[l]{Rocket Launch Ignition}}
& Item 66 & Up to T+0.175 s after launch, the hottest exhaust is concentrated mainly \_\_\_\_ . & Quant Est - Rel (B) & 1   & \gold{0.68} & \green{0.37} & -0.47 & 0.97 \\
& Item 67 & How does the exhaust plume change right after ignition? & Pattern Rec - Trnd & 0.6 & \gold{0.85} & \red{0.08} & 2.11  & 0.50 \\
\midrule

\multirow[c]{5}{2.7cm}{\makecell[l]{Car Aerodynamics}}
& Item 68 & After the air reaches the front of the car, it \_\_\_\_ . & Pattern Rec - Trnd & 1   & \green{0.95} & \gold{0.21} & 1.44  & 1.27 \\
& Item 69 & \textcolor{gray}{Which statement most accurately describes the vorticity magnitudes of the airflow?} & Pattern Rec - Trnd & 1   & \red{0.23} & \red{-0.03} & -4.92 & 0.34 \\
& Item 70 & Which region of the car shows the highest kinematic pressure? & Quant Est - Rel (B) & 1   & \green{0.92} & \gold{0.20} & 1.63  & 0.91 \\
& Item 71 & Over the course of the animation, what is the approximate difference between the maximum and minimum kinematic pressure on the windshield? & Quant Est - Rel (Q) & 1   & \red{0.27} & \gold{0.14} & -3.85 & 0.43 \\
\bottomrule

\end{tabularx}
}
\end{table*}

\vspace{-0.05in}
\subsection{Pilot Test}

Before conducting the large-scale tryout test, we executed a small quantitative pilot to evaluate feasibility and to triage items for revision. The objectives were to (1) verify that the assessment can be completed within the intended time window, (2) identify unclear stems/captions and ambiguous evidence, (3) screen for technical issues affecting visualizations, and (4) obtain preliminary item-level signals to prioritize revisions before large-sample calibration.

{\bf Participants.}\
We recruited 30 participants (19 females) from Prolific. Participants were residents of the United States or the United Kingdom, native English speakers, and aged 18 to 65 years. Participants' highest completed education was high school (16.7\%), technical or community college (20.0\%), a bachelor's degree (43.3\%), or a master's degree or higher (20.0\%). We required a Prolific approval rate of over 98 and self-reported normal or corrected-to-normal vision. To ensure the assessment was viewable as intended, participation required a laptop or desktop computer with a screen at least 1024 pixels wide; the study interface automatically disqualified attempts to access the study on mobile devices or at resolutions below the target.

{\bf Procedure.}\
Participants first read an introduction page explaining the purposes of SVLAT and providing test instructions. The assessment included 59 items plus three attention-check questions. We randomized the order in which visualization techniques were shown, and within each technique, we randomized the order of its associated items. Participants selected the best answer for each question and could choose {\em Skip} if they were unsure. After moving to the next question, they could not go back. We included only participants who answered at least two of the three attention-check questions correctly.
We logged item responses and per-item response times to diagnose items that impose disproportionate time burden or reading load. Because some SVLAT visualizations may include short animations, this pilot focused on verifying that such media can be presented reliably; we did not impose or record a specific ``watch'' behavior beyond standard survey interaction.

{\bf Scoring.}\
To account for the possibility of obtaining correct answers by random guessing in multiple-choice items, we also computed a correction-for-guessing score where skipped responses were not counted as incorrect answers. Let \(R\) denote the number of correctly answered items, \(W\) the number of incorrectly answered items, and \(k\) the number of response options per item. The corrected score is defined as \(S = R - \frac{W}{k-1}\) \cite{Diamond-Sage1973}. This formula subtracts the expected number of items that could be answered correctly by chance alone under random guessing, thereby providing a more conservative estimate of participants' actual performance.

{\bf Results and revision.}\
Overall, the pilot indicated that SVLAT is feasible within the intended 45-minute window: the median completion time for the 59-item assessment was 37.21 minutes. We observed no technical issues (all visualizations loaded and displayed as intended under the enforced device and resolution requirements). Performance was relatively high, with a median raw score of 48 (range: 37 to 56) and a median correction-for-guessing score of 44.33 (range: 26.67 to 55.00). Several items nonetheless showed ceiling effects (near-perfect accuracy) and/or unusually short response times, suggesting they could be answered with minimal engagement with the visualization or that their distractors were overly transparent. These pilot signals, therefore, motivated targeted refinements to increase difficulty without altering the intended construct.

We applied four revision strategies. 
First, we strengthened distractors in nine items (5, 9, 18, 19, 37, 51, 57, 62, 70) by making incorrect options more plausible and more closely matched in specificity and phrasing, reducing the likelihood that participants could succeed without reading the legend, comparing regions precisely, or tracking change over time. 
Second, for two image-based comparison items (14, 35), we updated the marked regions in the image to create closer, more informative contrasts, thereby reducing reliance on a single salient hotspot and increasing the need to extract quantitative or relative evidence from the encoding. 
Third, we revisited a subset of T/F items because pilot diagnostics indicated that some exhibited ceiling effects and unusually short response times, suggesting limited discrimination and greater susceptibility to guessing. To increase construct-relevant difficulty while maintaining low reading burden, we converted five items (16, 17, 23, 59, 69) to concise ``{\em Which statement most accurately describes ...}''-style multiple-choice formats, using short, tightly parallel alternatives that reflect common misinterpretations. 
Fourth, we redesigned four items (31, 41, 49, 61) mainly because the pilot indicated ambiguity or unintended shortcuts in the stem or evidence target, refining reference frames and evidence requirements so that success depends on interpreting the visualization rather than on background knowledge or coarse heuristics. 
In total, 20 items were revised across the above categories. 
Finally, we removed eight items (6, 24, 29, 30, 34, 38, 39, 58) that were least effective under the pilot diagnostics.
As a result, as Table~\ref{tab:item-list} shows, the revised SVLAT instrument comprises 18 visualizations and 51 items, improving alignment with the time budget and increasing construct-relevant difficulty prior to large-sample calibration. 


\vspace{-0.05in}
\subsection{Test Tryout}

We conducted a large-scale tryout to support item analysis and psychometric evaluation. The tryout was administered online via Qualtrics to participants recruited from Prolific, using a screen-out procedure similar to that in the pilot test.
The tryout followed the same procedure as the pilot test. Participants first read the study's purposes and instructions, then completed the assessment in Qualtrics. The assessment included 51 items plus three attention-check questions.

{\bf Participants.}\
A total of 500 participants were recruited. As in the pilot test, participants were required to complete the assessment on a device with a screen width of at least 1024 pixels. Eight participants were excluded for failing this requirement. We further excluded four random clickers who spent less than 10 seconds on more than 17 items (33\% of the assessment).
We also screened for insufficient-effort responding using two converging indicators:\ unusually short completion time and extensive skipping. Because the Tukey lower-fence rule \cite{Tukey-Nature1977} did not identify meaningful low-end outliers in completion time, we flagged respondents whose total completion time fell within the shortest 5\% of the sample. We additionally flagged respondents who skipped more than 25\% of items. Participants meeting both criteria were removed, resulting in the exclusion of three additional low-effort respondents.
We did not exclude participants based on low accuracy because accuracy is an intended outcome of SVLAT, and removing low scorers could bias score distributions and psychometric estimates. After these exclusions, the final sample comprised 485 participants (246 females). Participants' highest completed education was high school (17.7\%), technical or community college (15.3\%), a bachelor's degree (45.6\%), or a master's degree or higher (21.4\%).


{\bf Descriptive statistics.}\
We examined the raw scores, corrected scores, completion times, response times, and skip rates of the test takers. The total possible score on the test was 51. Raw scores ranged from 13 to 48 ($M=35.18$, $SD=7.07$). Corrected scores, adjusted using the correction-for-guessing formula, ranged from 2.33 to 47.33 ($M=30.17$, $SD=9.19$). On average, the corrected scores were 5.01 points lower than the raw scores. Total completion time ranged from 14.70 to 145.40 minutes ($M=37.33$, $SD=15.46$). Based on these completion times, 45 minutes is a reasonable estimated duration for completing the test for most participants. The overall skip rate across responses was 6.7\%.

\vspace{-0.05in}
\subsection{Item Analysis and Test Refinement}

Using the tryout response data, we analyzed item performance and refined the test using complementary CTT and Bayesian IRT evidence. 

\vspace{-0.05in}
\subsubsection{CTT}

For CTT item analysis, we examined two item-level indices:\ item difficulty and item discrimination. Item difficulty was indexed by the proportion of participants who answered each item correctly ($P$) \cite{thorndike1991measurement}, with larger values indicating easier items and smaller values indicating harder items. Item discrimination was assessed using the corrected item-total correlation ($r$), computed as the correlation between a focal item's score and the total test score with that item removed \cite{nunnally1978psychometric}. The corrected version was used instead of the uncorrected version to avoid part-whole inflation and provide a less biased estimate of how well an item differentiates between participants with lower and higher overall performance.
These CTT indices were used to identify items with undesirable response patterns, such as extremely low or high difficulty or weak discrimination. In general, items with very low corrected item-total correlations were considered candidates for revision or removal, especially when such items also showed problematic difficulty or qualitative evidence of ambiguity.



Table~\ref{tab:item-list} reports the CTT item statistics for the 51 tryout items. Overall, item difficulty $P$ ranged from \(0.18\) to \(0.97\), indicating that the tryout set covered a broad span from difficult to easy items. Using adopted CTT interpretive thresholds \cite{CTT-Thresholds}, 5 items were classified as hard ($P<0.50$), 30 as moderate ($0.50 \leq P < 0.85$), and 16 as easy ($P \geq 0.85$). The corrected item-total correlation $r$ ranged from \(-0.03\) to \(0.43\). Using corresponding discrimination thresholds, 15 items showed high discrimination ($r \geq 0.30$), 31 showed medium discrimination ($0.10 \leq r < 0.30$), and 5 showed low discrimination ($r<0.10$).

Item 69 was identified as a clear candidate for removal because it was very difficult and showed negative discrimination, indicating that it did not function as intended. Several items were flagged for further review based on their CTT profiles. Items 15, 49, 57, and 67 of low discrimination indicated limited ability to distinguish between lower- and higher-performing participants. These items were further reviewed to determine whether they provided sufficient information to justify retention. Overall, the CTT results indicated that most items performed adequately, while a small subset warranted closer inspection in the Bayesian IRT analysis and final item-selection step.

\vspace{-0.05in}
\subsubsection{Bayesian IRT}

We next conducted item analysis using a Bayesian 2PL IRT model, which is appropriate for dichotomously scored selected-response items such as the MCQ and T/F items in SVLAT \cite{demars2010item, burkner2021bayesian}. The Bayesian 2PL model estimates participant ability, item easiness, and item discrimination, allowing items to vary in both location and slope.

Let $Y_{ij}$ denote the response of participant $j$ to item $i$, where $Y_{ij}=1$ indicates a correct response and $Y_{ij}=0$ indicates an incorrect response. 
Skipped responses were treated as missing rather than incorrect. Thus, only observed responses contributed to model estimation, preserving the distinction between observed incorrect responses and omitted responses. 
The probability of a correct response was modeled as
\[
P(Y_{ij}=1 \mid \theta_j, a_i, e_i)
=
\frac{1}{1+\exp\left[-a_i(\theta_j + e_i)\right]},
\]
where $\theta_j$ is the latent ability (in our case, SciVis literacy) of participant $j$, $a_i$ is the discrimination parameter of item $i$, and $e_i$ is the easiness parameter of item $i$.
Under this parameterization, item difficulty is the {\em negative} of item easiness, that is, $b_i=-e_i$. Thus, larger easiness values correspond to easier items, whereas more negative easiness values correspond to harder items. Equivalently, items with larger positive difficulty values are harder. When a participant's ability equals an item's difficulty (i.e., $\theta_j=-e_i$), the model-implied probability of a correct response is 0.50. Item discrimination $a_i$ determines how sharply the probability of a correct response changes as ability increases; items with larger discrimination values better differentiate participants at nearby ability levels.

The Bayesian 2PL model was fit in \texttt{R} using the \texttt{brms} package with \texttt{CmdStanR}/\texttt{Stan} as the backend. To identify the latent scale, the participant-level standard deviation of ability was fixed to 1. We used weakly informative priors informed by prior Bayesian IRT practice \cite{burkner2021bayesian} and the CALVI modeling framework \cite{ge2023calvi}: normal$(0,1)$ priors for the fixed effects of item easiness and log-discrimination, normal$(0,1)$ priors for the item-level standard deviations of easiness and log-discrimination, and an LKJ$(2)$ prior for the correlation structure of the item-level effects. The discrimination parameter was modeled on the log scale and exponentiated to ensure positivity.


For estimation, we ran four Markov chains with 20{,}000 iterations per chain, used 5{,}000 warmup iterations per chain, and thinned the retained draws by 5, yielding 12{,}000 total post-warmup draws. Convergence was evaluated using the potential scale reduction factor $\hat{R}$, bulk effective sample size, tail effective sample size, and trace-plot inspection. The minimum bulk effective sample size was 5{,}224, the minimum tail effective sample size was 8{,}103, and the maximum $\hat{R}$ value was 1.001, indicating satisfactory convergence. Posterior predictive checks were also inspected to evaluate overall model fit.

Posterior medians were used as point estimates for item easiness and item discrimination, and 66\% and 95\% credible intervals were used to quantify parameter uncertainty. Item review was guided by four complementary sources of Bayesian IRT evidence: (1) item easiness, to ensure that the retained item bank covered a broad range of proficiency; (2) item discrimination, to identify items that contributed little differentiation among participants; (3) credible intervals around item parameters, to flag unstable or weakly estimated items; and (4) visual inspection of ICCs, to identify items with flat, extreme, or otherwise undesirable response functions.

\begin{figure}[!t]
    \centering
    \includegraphics[width=0.485\textwidth]{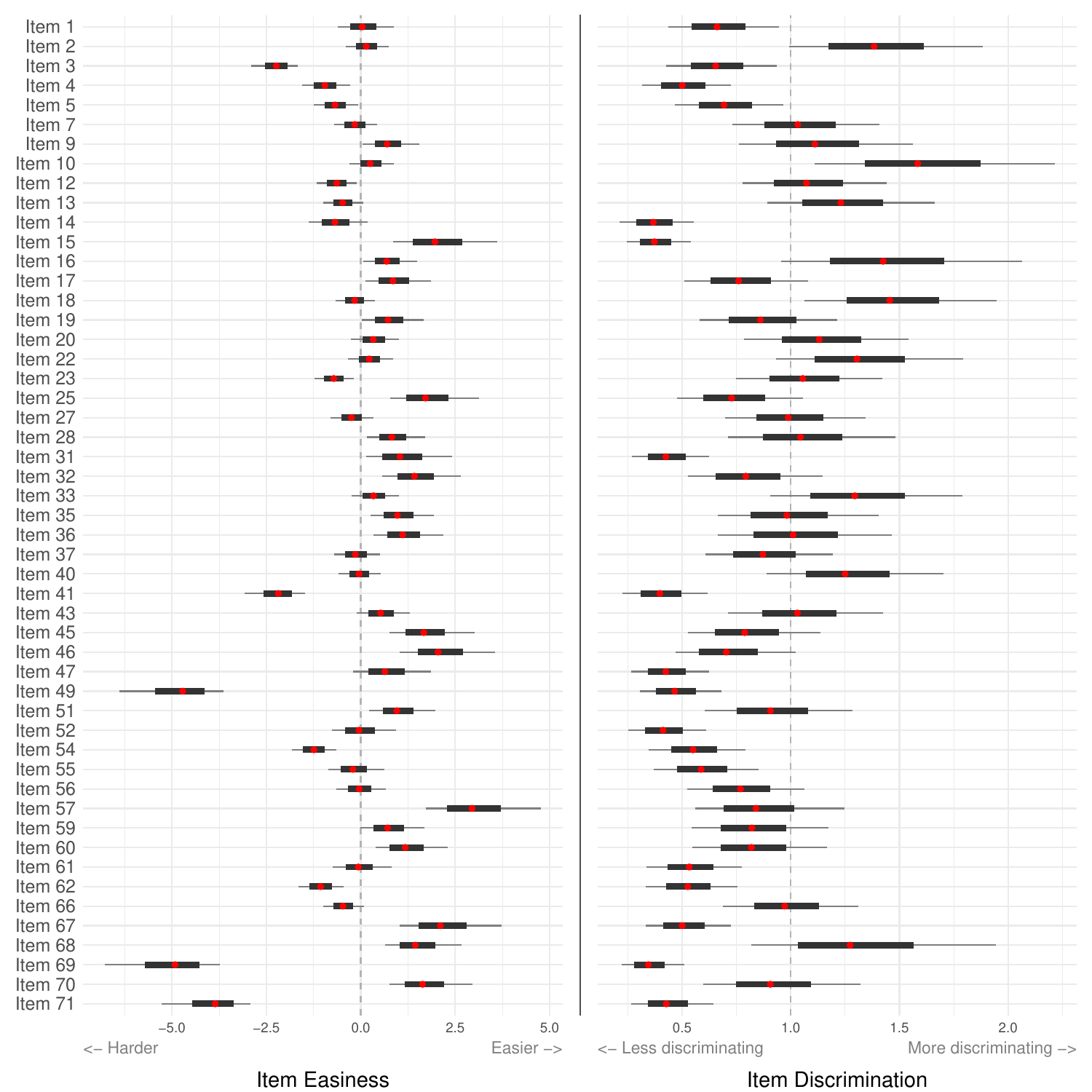}
    \vspace{-0.25in}
    \caption{IRT item parameter summary for the 51 SVLAT tryout items. Each dot shows the median easiness/discrimination estimate, with the lighter bars indicating the 95\% credible intervals and the darker bars indicating the 66\% credible intervals.}
    \label{fig:irt-item-params}
\end{figure}

Table \ref{tab:item-list} and Figure \ref{fig:irt-item-params} report the posterior median item easiness \(e\) and item discrimination \(a\) estimates for the 51 items. Posterior median easiness ranged from \(-4.92\) to \(2.95\), indicating broad coverage across the latent SciVis literacy scale. Posterior median discrimination ranged from \(0.34\) to \(1.58\), indicating substantial variation in the extent to which items differentiated participants. Following Baker's practical guidelines, item location values typically fall between \(-3\) and \(+3\), and discrimination values below \(0.50\) are weaker than those usually seen in well-maintained item pools \cite{IRT-Thresholds}. Under this interpretation, Items 49 (\(e = -4.72\), \(a = 0.47\)), 69 (\(e = -4.92\), \(a = 0.34\)), and 71 (\(e = -3.85\), \(a = 0.43\)) were notably problematic because they were substantially more difficult than the typical range and also showed weak discrimination. Items 14, 15, 31, 41, and 52 also showed weak discrimination, but their easiness estimates remained within a reasonable range. Overall, the Bayesian IRT results indicated that most items functioned within an acceptable range, while a smaller subset showed weak discrimination, extreme easiness, or both.

\vspace{-0.05in}
\subsubsection{Final Item Selection}

Final item selection was based on converging evidence across the CTT and Bayesian IRT analyses, rather than on any single statistic alone. Items were prioritized for removal when they showed consistently problematic performance across both frameworks, particularly when they combined extreme difficulty with weak discrimination.

Across the two analyses, Items 49 and 69 emerged as the clearest problematic cases. Both items were the two most difficult items in the item pool. They also showed low corrected item-total correlations in the CTT results and weak discrimination in the IRT results. Thus, both methods converged in indicating that these two items contributed less effectively to measurement quality than the remaining items.

Other items were flagged for review but showed less consistent evidence across methods. For example, Item 15 showed low discrimination but had an acceptable difficulty level in both analyses. Items 57 and 67 had low discrimination in the CTT analysis, but their weakness was less pronounced in the Bayesian IRT results. Conversely, Item 71 showed a problematic profile in the IRT analysis, but the evidence for removal was not as strong across both frameworks as it was for Items 49 and 69. Because the goal was to remove only items with the clearest and most consistent psychometric weaknesses, these items were retained for the present version.

\begin{figure}[!t]
    \centering
    \includegraphics[width=\columnwidth]{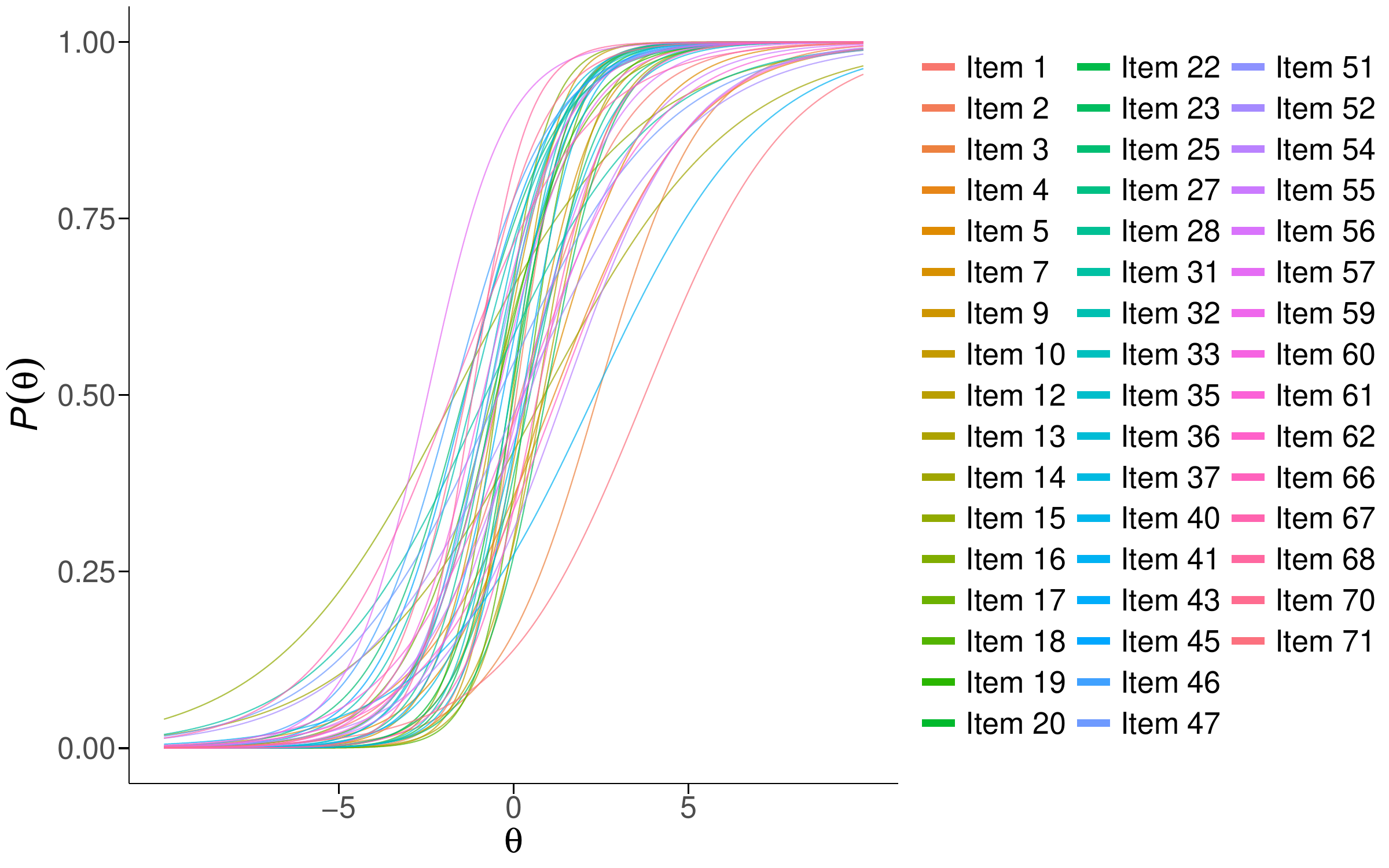}
    \vspace{-0.25in}
    \caption{ICCs for the final 49 SVLAT items under the Bayesian 2PL IRT model. Each curve shows the modeled probability of a correct response ($P(\theta)$) as a function of latent SciVis literacy ability ($\theta$).
    }
    \label{fig:svlat-icc}
\end{figure}

Based on this combined review, we removed two items (49, 69) from the tryout set. The final SVLAT consists of 49 items: 41 four-option MCQs, 1 three-option MCQ, and 7 T/F questions. The average CVR across items was 0.79, indicating strong expert agreement on the essentiality of the retained items.
Figure~\ref{fig:svlat-icc} shows the corresponding ICCs under the refit Bayesian 2PL model.

\vspace{-0.05in}
\subsection{Reliability Evaluation}

The reliability of SVLAT was assessed by examining internal consistency using 
McDonald's $\omega_t$ \cite{mcdonald2013test} and Cronbach's $\alpha$ \cite{cronbach-1951}. For the final 49-item SVLAT, $\omega_t = 0.82$, suggesting a high level of internal consistency. In addition, $\alpha = 0.81$, indicating good reliability and exceeding the commonly used threshold of 0.7. We report $\omega_t$ alongside $\alpha$ because it is often regarded as a more informative reliability coefficient when item loadings on the underlying construct may vary. The close correspondence between these two coefficients further confirms SVLAT's reliability in measuring SciVis literacy.

%% file: 5_discussion.tex
This work extends the visualization literacy assessment literature into the SciVis domain. Prior assessment research has shown the value of standardized, psychometrically grounded instruments for general visualization literacy, misleading visualization literacy, and adaptive testing \cite{lee2016vlat, pandey2023mini, Boy-TVCG14, ge2023calvi, cui2023adaptive}. However, these efforts have largely focused on charts and other InfoVis-oriented representations rather than the broader range of visual forms used in SciVis. 

Our results suggest that SciVis literacy can likewise be operationalized as a measurable construct for general audiences. The final SVLAT provides a feasible fixed-form instrument spanning multiple techniques and task types, while maintaining high reliability, good internal consistency, and a broad spread of item difficulty and discrimination. Taken together, these results indicate that SciVis interpretation skills can be assessed in a standardized, psychometrically grounded manner rather than solely through ad hoc task sets or domain-specific evaluations.

The psychometric results are also meaningful in a practical sense. The final item set does not merely show high reliability; it also retains enough variation in item difficulty and discrimination to distinguish participants across different ability levels. This is important because an assessment that is highly reliable but overly easy, overly hard, or too homogeneous would have limited usefulness for comparing participants or evaluating interventions. 

Our refinement further shows why combining CTT and Bayesian IRT is valuable. CTT provided transparent evidence about observed item difficulty and discrimination, while Bayesian IRT clarified item behavior on a common latent scale and more explicitly identified items with extreme easiness or weak slopes. The convergence of these two perspectives made the final pruning decisions more defensible and resulted in a more balanced instrument.

{\bf Broader relevance and uses.}\
SVLAT has several potential uses in research and practice. First, it provides a common instrument for SciVis studies, thereby improving comparability across datasets, techniques, and participant populations. This aligns with earlier arguments that visualization literacy assessments are valuable not only for research, but also for design and teaching, because they help characterize what audiences can interpret and where difficulties arise \cite{Boy-TVCG14}. Second, SVLAT can support educational and outreach settings by enabling more systematic evaluation of instructional materials, training activities, and public-facing scientific communication. In this sense, it can help move SciVis pedagogy toward the kind of concise, repeatable assessment workflows that prior adaptive visualization literacy work has argued are important for tracking ability and evaluating interventions efficiently \cite{cui2023adaptive}. Third, SVLAT may be useful beyond human subject studies as a benchmark for AI systems. Recent work has already used standardized visualization literacy tests to evaluate VLMs. It has been shown that current models still struggle with many visualization interpretation tasks, especially those that require more complex reasoning \cite{pandey2025benchmarking, Hong-TVCG25}. A SciVis-specific assessment, therefore, opens the door to more meaningful human-AI comparisons on scientific visualizations rather than relying only on InfoVis benchmarks.

{\bf Limitations.}\
Several limitations should be considered when interpreting the current results. First, the fixed-form assessment was designed to cover a broad range of visualization techniques and task types within an intended 45-minute window. While this breadth is a strength, the test is still fairly long, which may limit its practicality in some settings. One potential next step is therefore to explore shortened or adaptive variants, as prior work has shown that adaptive visualization literacy assessments can substantially reduce testing burden while preserving measurement quality \cite{pandey2023mini, cui2023adaptive}. Second, the current instrument emphasizes interpretation from static figures, short animations, and captions, with only limited coverage of interaction-dependent SciVis tasks. As a result, it does not yet capture important real-world activities such as navigation, slicing, parameter adjustment, or iterative exploratory analysis. Relatedly, uncertainty-focused visualizations are underrepresented, even though uncertainty is central to many scientific workflows.

%% file: 6_conclusion.tex
We introduced SVLAT, a scientific visualization literacy assessment test that measures how well general audiences read, understand, and interpret scientific visualizations and illustrations. To the best of our knowledge, SVLAT is among the first psychometrically grounded instruments designed specifically for this purpose. 

Through a staged development pipeline involving construct definition, blueprint construction, item generation, content validity, pilot test, large-scale tryout, item analysis, and reliability evaluation, we refined the instrument into a final 49-item assessment spanning 18 scientific visualizations and illustrations across eight visualization techniques and 11 tasks.
%
The instrument showed encouraging psychometric properties, including strong content validity from expert review (mean CVR = 0.79) and good internal consistency in the tryout sample (McDonald's $\omega_t = 0.82$, Cronbach's $\alpha = 0.81$).

Our results show that SVLAT can be administered within a practical online testing setting and provides a reliable measure of SciVis literacy. The combined use of CTT and Bayesian IRT supported principled item analysis and refinement, yielding an instrument with good internal consistency and a broad range of item difficulty and discrimination. Beyond its immediate value as an assessment tool, SVLAT provides a foundation for future research on SciVis literacy, educational evaluation, and benchmarking human and AI performance on SciVis interpretation tasks. We hope this work contributes to a broader measurement ecosystem for understanding and improving how people engage with scientific visualizations.

%% file: 7_suppl.tex

Supplementary materials are available at \url{https://osf.io/hr3nw/}
 under a CC BY-NC-SA 4.0 license. The repository includes the final SVLAT test PDF/item bank, all visualizations used in the assessment, analysis scripts, figure credits and source links, and ParaView state files/scripts for the in-house renderings. These materials support transparency, reproducibility, and proper attribution of the assessment and its visualization assets. 

%% file: 8_credit.tex
Figure~\ref{fig:visualizations} credits:\ 
(1) Terrain Map of the Roberts Mountains, Nevada---Image adapted from USGS National Geologic Map Database.
(2) Hurricane Laura Forecast---Image adapted from NOAA National Hurricane Center.
(3) Sea Surface Temperature Trend---Image adapted from NOAA Coral Reef Watch.
(4*) Wind Map of the United States---Animation from hint.fm.
(5) Molecule Structure---Image from free3d.com.
(6) Closed Seifert Surface---Image adapted from Jarke J. van Wijk (Eindhoven University of Technology, Netherlands).
(7) Human Body Organs---Data from The Cancer Imaging Archive. Image created using ParaView.
(8) Human Brain Fiber Bundles---Image adapted from Zeynep Saygin / MIT Koch Institute.
(9) Multiscale DNA Structure---Image from Cleveland Clinic.
(10) SARS-CoV-2 Virus---Image from Hangping Yao et al., Cell 183(3):730-738, 2020.
(11) Dragonfly Anatomy---Image adapted from M. A. Broussard @ Wikimedia Commons.
(12) Carp Skeleton---Data from Michael Scheuring (University of Erlangen, Germany). Image created using ParaView.
(13*) Black Hole Accretion---Animation adapted from Yan-Fei Jiang (Flatiron Institute) and Patrick Moran (NASA Ames Research Center).
(14*) Core-Collapse Supernova---Data from John M. Blondin (North Carolina State University) and Anthony Mezzacappa (Oak Ridge National Laboratory). Animation created using ParaView.
(15) Electric Field of a Capacitor---Image created using Python code from scipython.com.
(16) Velocity Field of a Tornado---Data produced using C code from Roger A. Crawfis (The Ohio State University). Image created using ParaView.
(17*) PC Airflow---Animation from Pikoandniki @ Youtube.
(18*) Rocket Launch Ignition---Animation adapted from Michael Barad and Tim Sandstrom (NASA Ames Research Center).
(19*) Car Aerodynamics---Animation adapted from CFDSupport @ Youtube.

%% file: 9_appendix.tex
\newpage
\clearpage

\setcounter{section}{0}
\setcounter{figure}{0}
\setcounter{table}{0}
\setcounter{page}{1}


\section{CTT and IRT Item Analysis}

\begin{figure}[H]
    \centering
    \includegraphics[width=\columnwidth]{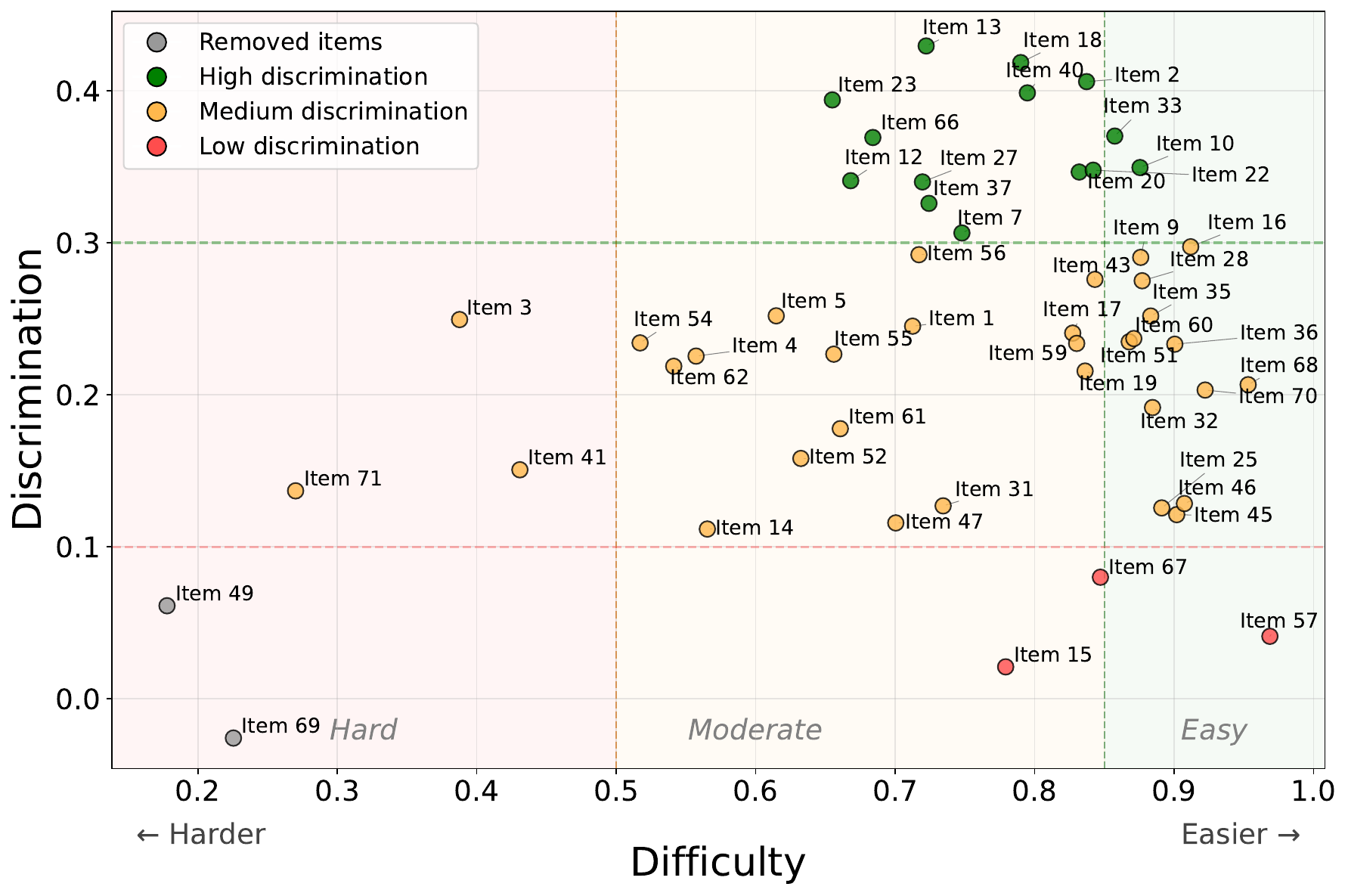}
    \vspace{-0.25in}
    \caption{CTT item difficulty and discrimination for the 51 SVLAT tryout items. Items 49 and 69 are removed from the final SVLAT version.}
    \label{fig:svlat-ctt}
\end{figure}

\begin{figure}[H]
    \centering
    \includegraphics[width=\columnwidth]{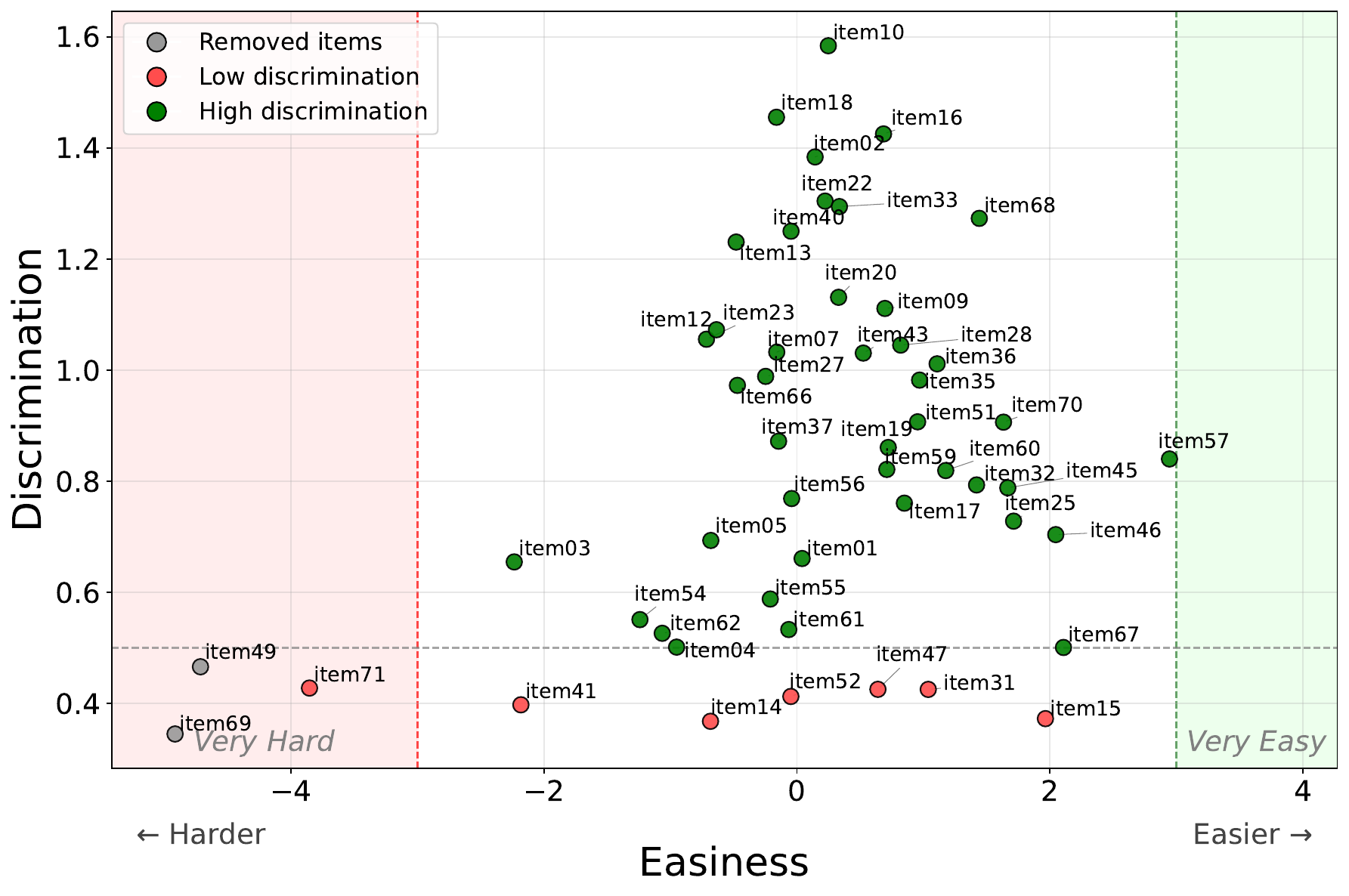}
    \vspace{-0.25in}
    \caption{Posterior median Bayesian IRT item easiness and discrimination for the 51 SVLAT tryout items. Items 49 and 69 are removed from the final SVLAT version.}
    \label{fig:svlat-irt}
\end{figure}

Figures \ref{fig:svlat-ctt} and \ref{fig:svlat-irt} show supplementary scatterplots for CTT and IRT item analysis, revealing several patterns across the two methods. First, easy items were not necessarily weak items: many items (e.g., Items 16, 68, and 70) in the easy range still showed medium or high discrimination in both CTT and Bayesian IRT, indicating that higher easiness alone was not grounds for removal. Second, the strongest discriminating items were concentrated mainly in the moderate-to-easy portion of the pool. In contrast, the hardest items tended to discriminate less well, especially Items 49 and 69 and, to a lesser extent, Item 71. This pattern helps explain why the assessment provides its strongest measurement around the middle of the ability range rather than at the extremes (see Figure \ref{fig:svlat-test-info-function} in Appendix \ref{app:timp}). Finally, the plots show that removing Items 49 and 69 improves the overall balance of the instrument without substantially narrowing the spread of remaining item difficulty, supporting the claim that the final 49-item SVLAT preserves broad coverage while improving psychometric quality.

\vspace{-0.05in}
\section{Bayesian IRT Results by Visualization Technique}

\begin{figure}[H]
    \centering
    \includegraphics[width=\columnwidth]{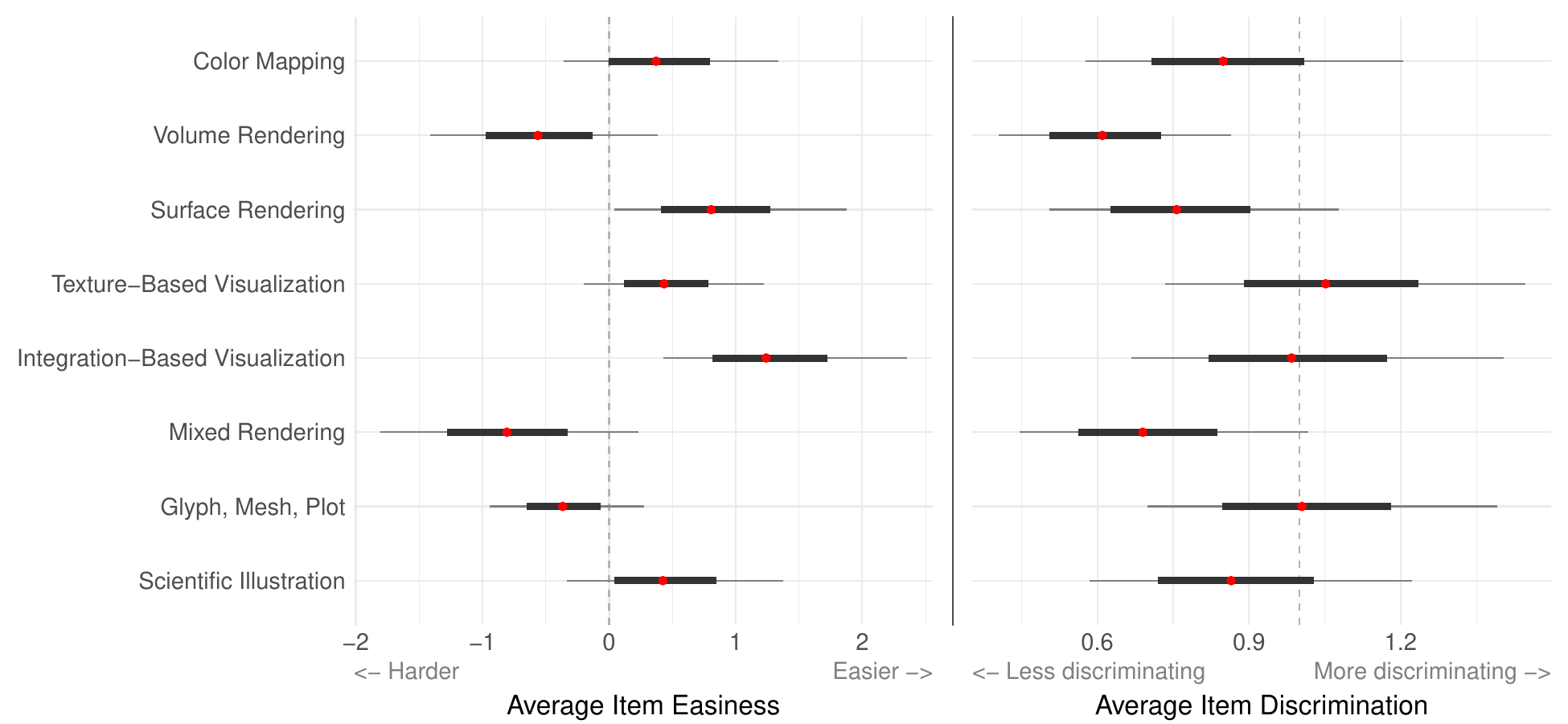}
    \vspace{-0.25in}
    \caption{IRT average item parameter summary for each visualization technique for the 51 SVLAT tryout items. Each dot shows the median easiness/discrimination estimate, with the lighter bars indicating the 95\% credible intervals and the darker bars indicating the 66\% credible intervals.}
    \label{fig:svlat-irt-by-technique}
\end{figure}

Figure~\ref{fig:svlat-irt-by-technique} suggests that the average item parameters varied meaningfully across visualization techniques. In the easiness panel, {\em Integration-Based Visualization} appears to have the highest average easiness, indicating that these items were, on average, easier for participants. In contrast, {\em Mixed Rendering} items appear to be the hardest on average. {\em Surface Rendering} and {\em Texture-Based Visualization} also trend toward the easier side, while {\em Glyph, Mesh, Plot} and {\em Volume Rendering} fall below zero, indicating relatively harder items overall. In the discrimination panel, {\em Texture-Based Visualization} and {\em Glyph, Mesh, Plot} show the highest average discrimination, suggesting stronger differentiation among participants, whereas {\em Volume Rendering} and {\em Mixed Rendering} show lower average discrimination.

\vspace{-0.05in}
\section{Relationship between Raw Score and Latent Ability}

\begin{figure}[H]
    \centering
    \includegraphics[width=0.8\columnwidth]{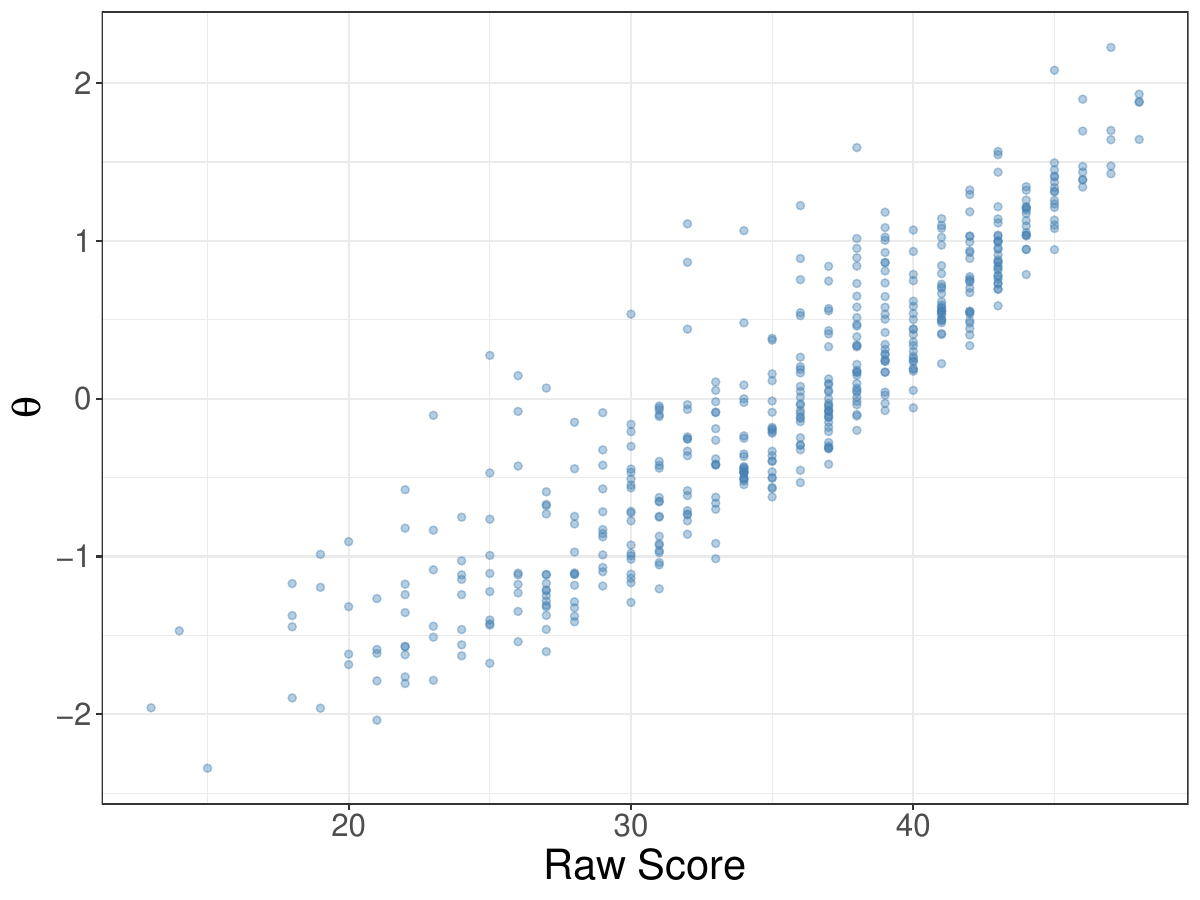}
    \vspace{-0.125in}
    \caption{Relationship between participants' raw score and their estimated latent SciVis literacy ability ($\theta$) under the Bayesian 2PL IRT model. Each point represents one participant in the SVLAT tryout sample.}
    \label{fig:svlat-ability-and-raw-score}
\end{figure}

Figure~\ref{fig:svlat-ability-and-raw-score} shows a strong positive relationship between participants' raw score and their estimated latent SciVis literacy ability under the Bayesian 2PL IRT model. Participants with higher raw scores generally received higher \(\theta\) estimates, indicating that the latent ability estimates were well aligned with observed overall test performance. At the same time, the vertical spread of points at a given raw score shows that participants with the same raw total could still differ somewhat in their estimated ability, reflecting the IRT model's use of item parameters rather than raw score alone. This pattern suggests that raw scores provide a useful overall summary of performance, while latent ability estimates offer a more refined representation of participant proficiency.

\vspace{-0.05in}
\section{Relationship between Skip Rate and Latent Ability}

\begin{figure}[H]
    \centering
    \includegraphics[width=0.8\columnwidth]{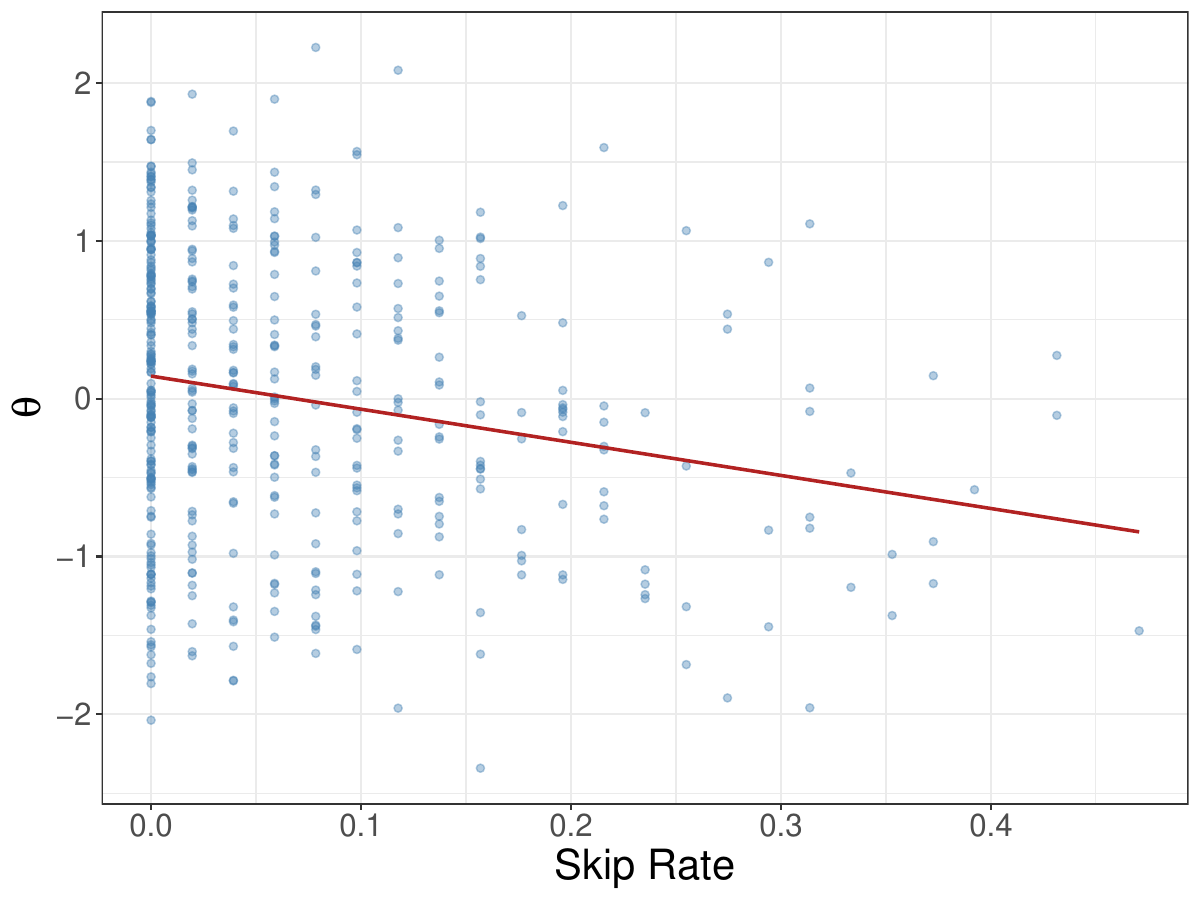}
    \vspace{-0.125in}
    \caption{Relationship between participants' skip rate and their estimated latent SciVis literacy ability ($\theta$) under the Bayesian 2PL IRT model. Each point represents one participant in the SVLAT tryout sample.}
    \label{fig:svlat-ability-and-skiprate}
\end{figure}

Figure~\ref{fig:svlat-ability-and-skiprate} shows a modest negative association between participants' skip rate and their estimated latent SciVis literacy ability under the Bayesian 2PL IRT model. Participants who skipped more items tended to have lower \(\theta\) estimates on average, and this relationship was statistically significant (\(r=-0.21, p<0.001\)). At the same time, the substantial vertical spread of points, especially near very low skip rates, indicates that skip behavior alone did not determine ability.

\vspace{-0.05in}
\section{Test Information and Measurement Precision}
\label{app:timp}

\begin{figure}[H]
    \centering
    \includegraphics[width=0.75\columnwidth]{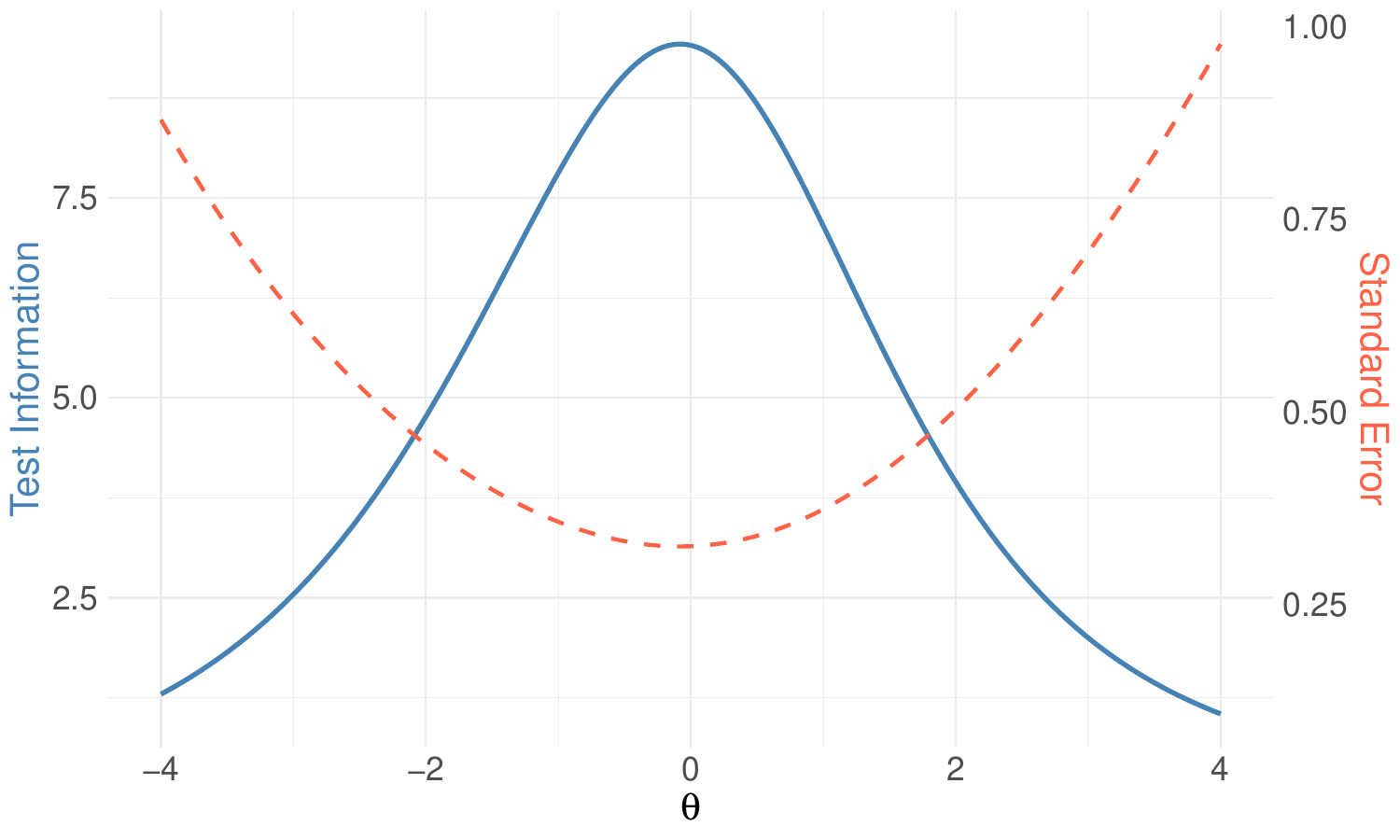}
    \vspace{-0.1in}
    \caption{Test information function (solid blue) and the corresponding standard error of measurement (dashed red) for the final SVLAT under the Bayesian 2PL IRT model.}
    \label{fig:svlat-test-info-function}
\end{figure}

Figure~\ref{fig:svlat-test-info-function} shows that the final SVLAT provides its greatest measurement precision around the middle of the latent SciVis literacy scale, where the test information function reaches its peak, and the standard error of measurement is lowest. Information declines and standard error increases toward both the lower and higher ends of latent SciVis literacy ability ($\theta$), indicating that the assessment is less precise for participants at the extremes of ability. This pattern is consistent with the item pool composition, in which many retained items are concentrated in the moderate-to-easy range and therefore provide the most information for participants near the center of the latent trait distribution.

